\documentclass[lettersize,journal]{IEEEtran}
\usepackage{amsmath,amsfonts,amssymb}
\usepackage{array}
\usepackage[caption=false,font=footnotesize]{subfig}
\usepackage{textcomp}
\usepackage{stfloats}
\usepackage{url}
\usepackage{verbatim}
\usepackage{graphicx}
\usepackage{cite}
\usepackage[hidelinks]{hyperref}
\usepackage[linesnumbered,ruled,vlined]{algorithm2e}

\usepackage[acronym]{glossaries}
\newacronym{LLM}{LLM}{large language model}
\newacronym{5G}{5G}{fifth generation}
\newacronym{6G}{6G}{sixth generation}
\newacronym{BS}{BS}{base station}
\newacronym{CoT}{CoT}{chain-of-thought}
\newacronym{ToT}{ToT}{tree-of-thought}
\newacronym{GoT}{GoT}{graph-of-thought}
\newacronym{ICL}{ICL}{in-context learning}
\newacronym{PCoT}{PCoT}{plan-based CoT}
\newacronym{MAE}{MAE}{mean absolute error}
\newacronym{M2M}{M2M}{machine-to-machine}
\newacronym{PE}{PE}{permutation entropy}
\newacronym{RMSE}{RMSE}{root mean square error}
\newacronym{WMA}{WMA}{weighted moving average}
\newacronym{SMA}{SMA}{simple moving average}
\newacronym{LSTM}{LSTM}{long short-term memory}
\newacronym{ARIMA}{ARIMA}{auto regressive integrated moving average}
\newacronym{AGCRN}{AGCRN}{adaptive graph convolutional recurrent network}
\newacronym{StemGNN}{StemGNN}{spectral temporal graph neural network}
\newacronym{SDGNet}{SDGNet}{spatiotemporal dynamic graph network}
\newacronym{API}{API}{application programming interface}
\newacronym{GPU}{GPU}{graphics processing unit}
\newacronym{TFT}{TFT}{temporal fusion transformer}
\newacronym{STDT-Net}{STDT-Net}{spatial-temporal downsampling transformer neural network}
\newacronym{DGC}{DGC}{dynamic graph convolution}
\newacronym{UAV}{UAV}{unmanned aerial vehicle}
\newacronym{DL}{DL}{deep learning}
\newacronym{ML}{ML}{machine learning}
\newacronym{GLU}{GLU}{gated linear unit}
\newacronym{NLP}{NLP}{natural language processing}
\newacronym{QoS}{QoS}{quality of service}
\newacronym{QoE}{QoE}{quality of experience}
\newacronym{AI}{AI}{artificial intelligence}
\newacronym{RSRP}{RSRP}{reference signal received power}

\glsdisablehyper
\usepackage{bm}
\usepackage{booktabs}
\usepackage[table]{xcolor}
\definecolor{myyellow}{RGB}{245,193,0}
\usepackage{makecell}
\usepackage{multirow}

\begin{document}

\title{Chain-of-Thought Reasoning Enhances In-Context Learning for LLM-Based Mobile Traffic Prediction}

\author{MohammadMahdi Ghadaksaz, Mohammad Farzanullah,~\IEEEmembership{Graduate Student Member,~IEEE}, Akram Bin Sediq, Ali Afana, Melike Erol-Kantarci, ~\IEEEmembership{Fellow,~IEEE}
\thanks{MohammadMahdi Ghadaksaz, Mohammad Farzanullah, and Melike Erol-Kantarci are with the School of Electrical Engineering and Computer Science, University of Ottawa, Ottawa, ON K1N 6N5, Canada (e-mail: mghad017@uottawa.ca; mfarz086@uottawa.ca; melike.erolkantarci@uottawa.ca).}
\thanks{Akram Bin Sediq and Ali Afana are with Ericsson, Ottawa, K2K 2V6, Canada (e-mail:
akram.bin.sediq@ericsson.com; ali.afana@ericsson.com)
}
}



\maketitle

\begin{abstract}
Accurate short-term mobile traffic prediction is important for proactive resource allocation and low-latency network management in \gls{5G} and \gls{6G}. While \glspl{LLM} can perform \gls{ICL} without task-specific retraining, naive \gls{ICL} prompting may suffer from numerical instability and limited temporal reasoning when traffic dynamics fluctuate rapidly. In this paper, we propose a \gls{CoT}-enabled \gls{LLM}-based mobile traffic prediction framework that operates in two phases: (i) an offline phase that constructs structured \gls{CoT} demonstrations by generating \emph{rationales} via a \gls{PCoT} pipeline (\emph{lecture}, \emph{plan}, and \emph{rationale}), and (ii) an online phase that performs close to real-time prediction by retrieving the most relevant demonstrations using a similarity policy that considers both the historical throughput pattern and its short-term changes. We evaluate the proposed framework using a real-world \gls{5G} measurement dataset that includes both driving and static scenarios across diverse applications. Our numerical results reveal that the proposed 2-shot \gls{CoT}-\gls{LLM} can improve \gls{MAE}, \gls{RMSE} and $R^2$-score by up to $14.88\%$, $15.03\%$, and $22.41\%$, respectively, compared to the 2-shot \gls{ICL}-\gls{LLM} and classical baselines. Furthermore, by optimizing the number of in-context examples, we achieve additional improvements of $4.58\%$, $5.70\%$, and $4.85\%$ in \gls{MAE}, \gls{RMSE}, and $R^2$-score, respectively. 
\end{abstract}

\begin{IEEEkeywords}
Large language models (LLMs), chain of thought, reasoning, mobile traffic prediction, 6G.
\end{IEEEkeywords}
\glsresetall[acronym]

\section{Introduction}
\IEEEPARstart{W}{ireless} mobile networks are experiencing unprecedented growth in traffic volume and variability due to the proliferation of data-intensive applications such as high-definition video streaming, cloud-based services, and interactive mobile platforms. Beyond \gls{5G}, emerging \gls{6G} networks will introduce new demands from physical \gls{AI}, agentic communications, extended reality, and \gls{M2M} interactions, creating more diverse and bursty traffic patterns that traditional prediction frameworks struggle to capture. Notably, \gls{AI}-driven and \gls{M2M} workloads are highly latency-sensitive, making accurate traffic prediction essential to proactively allocate resources and meet service-level agreements before demand surges occur \cite{Lykakis2025DataTrafficPrediction, Saad2020SixG, 11370176}. \par 
Accurate traffic prediction allows network operators to transition from reactive to proactive network management. By anticipating future traffic demands, \glspl{BS} and core network entities can perform intelligent resource allocation, congestion avoidance, and adaptive \gls{QoS} provisioning. Moreover, predictive traffic awareness plays a central role in energy-efficient networking, mobility management, and edge intelligence, where timely decisions must be made under strict latency constraints \cite{9112608, 9352246}. However, wireless traffic is inherently complex, as it is jointly affected by user mobility, radio channel variations, application-level behavior, and network configurations. These factors make mobile traffic prediction a challenging task, especially in realistic, large-scale deployments. \par 
Traditional traffic prediction methods, including statistical models and classical time-series techniques, often rely on strong assumptions about stationarity and linearity, which limit their effectiveness in highly dynamic wireless environments. While \gls{ML}-based approaches have improved prediction accuracy by capturing nonlinear dependencies, they typically require large labeled datasets, extensive offline training, and periodic retraining to remain effective when traffic characteristics shift \cite{10279008}. Such requirements reduce their practicality in scenarios where traffic patterns change frequently or differ across locations and applications. \par 
Recently, \glspl{LLM} have emerged as a flexible alternative for data-driven inference and prediction tasks beyond their original focus on natural language processing. A key advantage of \glspl{LLM} is their ability to perform \gls{ICL}, where a model can adapt to a new task by observing a small number of examples provided directly in the prompt, without updating model parameters \cite{Brown2020LanguageModelsFewShot}. This capability makes \glspl{LLM} particularly appealing for mobile traffic prediction, as it enables rapid adaptation across diverse scenarios and reduces the need for repeated retraining. Nevertheless, naive \gls{ICL} prompting may struggle with numerical stability and temporal reasoning, especially when the prediction task involves complex traffic dynamics. \par

In this context, reasoning has emerged as a key capability in advanced \gls{AI} systems. By decomposing complex problems into intermediate logical steps, reasoning-enabled models improve generalization and interpretability. For \glspl{LLM}, this capacity is tied to the quality of intermediate steps generated before the final answer. \Gls{CoT} prompting has been introduced as an effective mechanism to improve the reasoning behavior of \glspl{LLM} by explicitly guiding the model through intermediate inference steps \cite{NEURIPS2022_9d560961}. By structuring the prediction process and exposing latent reasoning paths, \gls{CoT} prompting enables \glspl{LLM} to better exploit historical trends and contextual signals embedded in mobile traffic data. This motivates the exploration of \gls{CoT}-enabled \glspl{LLM} for wireless traffic prediction and necessitates an evaluation of their performance, robustness, and practical deployment in real-world scenarios.

\subsection{Related Works}
\subsubsection{Mobile Traffic Prediction}
Mobile traffic prediction has been an active research topic over the past few years \cite{10.1007/s10922-016-9392-x, TianL21-1, Trinh2018MobileTrafficLSTM, Bai2020AdaptiveGraphCRN, 9673775}. In particular, the authors in \cite{10.1007/s10922-016-9392-x} propose a \gls{WMA}-based approach for mobile traffic prediction. Similarly, \cite{TianL21-1} develops an \gls{ARIMA}-based statistical model for traffic prediction, showing that \gls{ARIMA} can outperform several benchmark methods. In \cite{Trinh2018MobileTrafficLSTM}, the authors propose an \gls{LSTM}-based framework to model and predict traffic patterns, where multivariate time-series prediction is performed for both one-step and longer-term predictions to evaluate how far ahead accurate predictions can be achieved. The work in \cite{Bai2020AdaptiveGraphCRN} introduces an \gls{AGCRN} to capture fine-grained spatial and temporal correlations in traffic data, demonstrating the effectiveness of graph-based modeling. Furthermore, \cite{9673775} investigates an \gls{SDGNet} framework based on \gls{DGC} and \glspl{GLU} to predict traffic consumption over short-, medium-, and long-term horizons, where the results show lower prediction error compared to conventional models. \par

With the advancement of transformer architectures \cite{Vaswani2017Attention} and their strong ability to capture long-term dependencies, these models have also been applied to mobile traffic prediction \cite{10946853, 9919315, 10840287, 10114636}. In \cite{10946853}, the authors introduce a \gls{TFT}-based framework for wireless traffic prediction to support efficient network management and improve \gls{QoE}. In \cite{9919315}, a \gls{STDT-Net} approach is proposed to jointly exploit temporal, local spatial, and global spatial dependencies for traffic prediction. Similarly, the studies in \cite{10840287, 10114636} investigate spatio-temporal transformer architectures for cellular traffic prediction, further highlighting the potential of transformer-based models in this domain.
\\
\subsubsection{\glspl{LLM} in Wireless Communications}
Thanks to their strong capabilities and proven success in both academia and industry \cite{Zhou2024LLMTelecomSurvey}, \glspl{LLM} have recently gained increasing attention in the wireless communications community for a wide range of prediction and detection tasks \cite{Zhang2024LLMWirelessIntrusion, HanICL, Hu2025SelfRefinedTrafficLLM, Habib2025LLMIntentNetOpt}. One key advantage of \gls{LLM}-based approaches is that, unlike conventional transformer- or \gls{ML}-based methods, they do not require task-specific retraining. While initial pre-training of \glspl{LLM} depends on massive corpora available in the language domain, wireless datasets—containing signals, channel measurements, or traffic traces—remain limited in public availability and costly to collect. Consequently, parameter-free adaptation through \gls{ICL} is particularly attractive for wireless tasks, where re-training on every scenario is impractical.

In \cite{Zhang2024LLMWirelessIntrusion}, the authors propose an \gls{LLM}-based intrusion detection framework using \gls{ICL}, and their results demonstrate acceptable detection accuracy. Similarly, the authors in \cite{HanICL} 
introduce an \gls{LLM}-based mobile traffic prediction framework with a two-stage \gls{ICL} example selection strategy, achieving low prediction error. In \cite{Hu2025SelfRefinedTrafficLLM}, a self-refined \gls{LLM} is designed to iteratively correct inaccurate predictions through a three-step process, where hourly traffic is predicted using random \gls{ICL} example selection. Moreover, the authors in \cite{Habib2025LLMIntentNetOpt} apply lightweight \glspl{LLM} to intent-processing tasks, demonstrating improved network throughput and efficiency. 

Several recent studies also investigate \gls{LLM} fine-tuning or training from scratch for time-series prediction tasks \cite{Cao2024TEMPO, Chang2025LLM4TS}. In \cite{Cao2024TEMPO}, the authors employ an interpretable, prompt-tuning-based generative transformer to learn time-series representations. Furthermore, \cite{Chang2025LLM4TS} proposes two fine-tuning strategies to better adapt \glspl{LLM} to the characteristics of time-series data, showing that fine-tuning can outperform \gls{ICL}-based methods and other prompt-engineering approaches in certain scenarios. \par 
Although \gls{CoT} prompting is still in its early stages of adoption in wireless communications, it has already been explored in several recent studies \cite{Huang2025ReasoningAI6G, wang2025chain}. In \cite{Huang2025ReasoningAI6G}, the authors apply \gls{CoT} to reason about the causes of performance degradation in \gls{6G} networks. Similarly, \cite{wang2025chain} investigates multiple \gls{CoT} strategies, where Auto-\gls{CoT} is employed for \gls{UAV} location and power allocation optimization. The results demonstrate that \gls{CoT}-based methods outperform their non-\gls{CoT} counterparts.

\subsection{Motivations \& Contributions}
Most existing research on wireless traffic prediction has largely overlooked the potential of \glspl{LLM}. This observation is evident in several prior works \cite{10.1007/s10922-016-9392-x, TianL21-1, Trinh2018MobileTrafficLSTM, Bai2020AdaptiveGraphCRN, 9673775, 10946853, 9919315, 10840287, 10114636}. In particular, studies such as \cite{10.1007/s10922-016-9392-x, TianL21-1} rely on simple statistical methods, which often struggle to capture complex traffic dynamics. Other works (i.e., \cite{Trinh2018MobileTrafficLSTM, Bai2020AdaptiveGraphCRN, 9673775, 10946853, 9919315, 10840287, 10114636}) adopt data-driven approaches that depend heavily on large training datasets, which are not always available, and whose training processes can be computationally expensive and time-consuming. On the other hand, although the authors in \cite{HanICL, Hu2025SelfRefinedTrafficLLM} have considered \glspl{LLM} for traffic prediction, they have been limited to \gls{ICL}, whereas the potential of \gls{CoT} prompting has been unexplored. Furthermore, several studies, such as \cite{Cao2024TEMPO, Chang2025LLM4TS}, employ fine-tuned \glspl{LLM} for time-series prediction. However, mobile traffic patterns evolve rapidly due to changes in user behavior, mobility, and application usage, which would require frequent fine-tuning and introduce additional computational overhead, complicating practical deployment in dynamic network environments. Moreover, fine-tuning still relies on the availability of task- and scenario-specific data, which may not always be readily available for all traffic conditions or deployment scenarios.  
In contrast, prompt-based approaches, such as \gls{ICL} and \gls{CoT}, do not require any model tuning, making them more flexible and easier to deploy in dynamic traffic environments. Lastly, although the works in \cite{Huang2025ReasoningAI6G, wang2025chain} employ \gls{CoT}-based solutions, they do not apply this approach to mobile traffic prediction. To the best of our knowledge, this is the first time \gls{CoT}-enabled \gls{LLM}-based mobile traffic prediction using a structured example selection has been explored to enhance performance using a real-world \gls{5G} dataset.
\par The main contributions of this work can be highlighted as follows:
\begin{enumerate}
	\item We propose a novel \gls{CoT}-enabled \gls{LLM}-based mobile traffic prediction framework that consists of an offline prompt construction phase and an online traffic prediction phase. During the offline phase, \emph{rationales} (i.e., step-by-step guidance toward accurate traffic prediction) are generated from historical traffic data using a three-step \emph{rationale} generation process, which serves as long-term memory to enhance prediction performance. Subsequently, during the online phase, the most similar examples are selected according to a specific selection policy, further improving the accuracy of traffic prediction.
	\item We evaluate the impact of the number of examples on the performance of the \gls{CoT}-\gls{LLM} for the traffic prediction task and analyze how the example selection policy affects the results. In addition, we compare the performance of \gls{CoT} with standard \gls{ICL} prompting and examine the stability of both approaches under varying numbers of examples and traffic conditions.
	\item We conduct an extensive evaluation across multiple open-weight \glspl{LLM}. In particular, since many state-of-the-art \glspl{LLM} are not publicly released, we show that the proposed \gls{CoT}-\gls{LLM}-based mobile traffic prediction framework can be implemented using open-weight models while achieving performance comparable to, or even exceeding, that of closed-weight models. By leveraging open-weight models,  \glspl{LLM} can be deployed  locally without relying on external \gls{API} services, thereby significantly reducing inference latency and enabling close to real-time traffic prediction. 
\end{enumerate}
\par We demonstrate the effectiveness of the proposed solutions using a real-world \gls{5G} dataset that includes diverse practical scenarios. Our numerical results validate the effectiveness of the proposed framework. In particular, the 2-shot \gls{CoT}-\gls{LLM} achieves improvements of up to \(14.88\%\), \(15.03\%\), and \(22.41\%\) in \gls{MAE}, \gls{RMSE}, and \(R^2\)-score, respectively, compared to the 2-shot \gls{ICL}-\gls{LLM} and classical baseline methods. Furthermore, we show that optimizing the number of in-context examples yields additional gains of \(4.58\%\), \(5.70\%\), and \(4.85\%\) in \gls{MAE}, \gls{RMSE}, and \(R^2\)-score, respectively, highlighting the importance of example selection. Finally, evaluations across multiple open-weight \glspl{LLM} demonstrate that locally deployable models can achieve performance comparable to, or exceeding, closed-weight alternatives, enabling practical low-latency traffic prediction without reliance on external \glspl{API}.


\subsection{Organization}
The remainder of this paper is organized as follows: Section \ref{sec:problem} presents the discussed problem followed by the methodology in Section \ref{sec:methodology}.
Section \ref{sec:results} provides the numerical results and analysis, and finally Section \ref{sec:conc} concludes the paper with conclusions and future works.  \par 
\section{PROBLEM DESCRIPTION}
\label{sec:problem}
In this section, we formulate the \gls{CoT}-enabled \gls{LLM}-based downlink mobile traffic prediction problem for a generic traffic measurement dataset that contains downlink throughput observations along with a set of network-related contextual features. \par
Let us denote the downlink throughput and the $k^{\text{th}}$ contextual feature at time step $t$ of measurement by $\gamma^{(t)}$ and $c_k^{(t)}$, respectively. A raw dataset $\tilde{\mathcal{D}}$ with $H$ seconds of traffic measurement is then represented as:
\begin{equation}
    \tilde{\mathcal{D}} = \left\{\left(\gamma^{(t)}, \mathbf{c}^{(t)}\right)\right\}_{t=1}^H,
\end{equation}
where $\mathbf{c}^{(t)} = \left[c_1^{(t)}, \cdots c_K^{(t)}\right] \in \mathcal{X} ^{K}$ ($\mathcal{X}$ denotes the mixed type data) collects the $K$ contextual features at time step $t$. Due to prompt-size limitations, we use a historical window of length $W$ seconds as the input to the prediction model. Subsequently, we define the historical throughput vector $\bm{\Gamma}^{(t)}$ at time step $t$ as:
\begin{equation}
\label{eq:past_throughput}
    \bm{\Gamma}^{(t)} = \left[\gamma^{(t)}, \gamma^{(t-1)}\cdots,\gamma^{(t-W+1)} \right] \in \mathbb{R}^{W}.
\end{equation}

Likewise, the historical contextual information matrix $\mathbf{C}^{(t)}$, including $K$ contextual features, at time step $t$ can be written as follows:
\begin{equation}
\label{eq:past_info}
\mathbf{C}^{(t)}=
\begin{bmatrix}
c_{1}^{(t)} & c_{2}^{(t)} & \cdots & c_{K}^{(t)} \\
c_{1}^{(t-1)} & c_{2}^{(t-1)} & \cdots & c_{K}^{(t-1)} \\
\vdots & \vdots & \ddots & \vdots \\
c_{1}^{(t-W+1)} & c_{2}^{(t-W+1)} & \cdots & c_{K}^{(t-W+1)}
\end{bmatrix} \in \mathcal{X}^{W \times K}.
\end{equation}

In this case, we focus on one time step traffic prediction task, expressed as:
\begin{equation}
    y^{(t)}=\gamma^{(t+1)},
\end{equation}
where using a pre-trained \gls{LLM} with fixed parameters, the model is asked to output the subsequent downlink throughput and a \emph{rationale} for its respond, formulated as:
\begin{equation}
\label{eq:zero-shot}
\left[\hat{y}^{(t)}, r^{(t)}\right] = f\left(\mathbf{C}^{(t)}, \bm{\Gamma}^{(t)} \vert\bm{\Theta}\right),
\end{equation}
where $\hat{y}^{(t)}$ denotes the predicted downlink throughput, $r^{(t)}$ is the \emph{rationale}, $f\left(\cdot\right)$ represents the \gls{LLM}, and $\bm{\Theta}$ is the model's parameters. \par 
Equation (\ref{eq:zero-shot}) is commonly referred to as zero-shot \gls{CoT}-\gls{LLM} prediction \cite{NEURIPS2022_8bb0d291}. To improve the performance of zero-shot prediction, we include several known examples, each paired with a \emph{rationale}, in the prompt. This forms an extension of zero-shot \gls{CoT}, which is known as few-shot \gls{CoT} \cite{NEURIPS2022_9d560961}. The incorporation of the known examples enables the \glspl{LLM} to solve new tasks without being fine-tuned or re-trained \cite{dong2024survey}. \par 
The \gls{CoT} prompting differs from \gls{ICL} prompting in that, in \gls{ICL}, the \gls{LLM} is only provided with examples, whereas in \gls{CoT} approaches, the examples are accompanied by an explicit \emph{rationale} that guides the model through the intermediate reasoning steps before producing the final prediction. Moreover, the model is explicitly instructed to \emph{think step-by-step} before generating the output, which encourages the \gls{LLM} to produce intermediate reasoning steps rather than jumping directly to a final answer — a technique shown to improve complex reasoning performance compared to standard \gls{ICL} prompting \cite{NEURIPS2022_8bb0d291, NEURIPS2022_9d560961}. Nonetheless, in both approaches, the selection and number of examples are important and can significantly impact the overall performance. \par 
To clearly separate the data used to construct the \gls{CoT} examples from the data used for evaluation, we assume that the training examples are drawn from a processed training dataset $\mathcal{D}_{\text{train}}$, whereas the test samples are taken out from a separate processed test dataset $\mathcal{D}_{\text{test}}$, with no overlap between the two datasets. Each example in the processed training set consists of a historical downlink throughput sequence and the associated contextual information, denoted by $\bm{\Gamma}_{\text{train}}^{(n)}$ and $\mathbf{C}_{\text{train}}^{(n)}$, respectively, along with a corresponding \emph{rationale} $r_{\text{train}}^{(n)}$, and the ground-truth traffic at the next time step $y_{\text{train}}^{(n)}$. This training example can be written as:
\begin{equation}
\label{eq:example}
    \mathbf{E}_{\text{train}}^{(n)} = \left[\left[\mathbf{C}_{\text{train}}^{(n)}, \bm{\Gamma}_{\text{train}}^{(n)}\right], r_{\text{train}}^{(n)}, y_{\text{train}}^{(n)}\right],
\end{equation}
\begin{figure*}[t]
    \centering

    \subfloat[\label{fig:myfig:a}]{
        \includegraphics[trim=16 80 19 106,clip, width=0.48\linewidth]{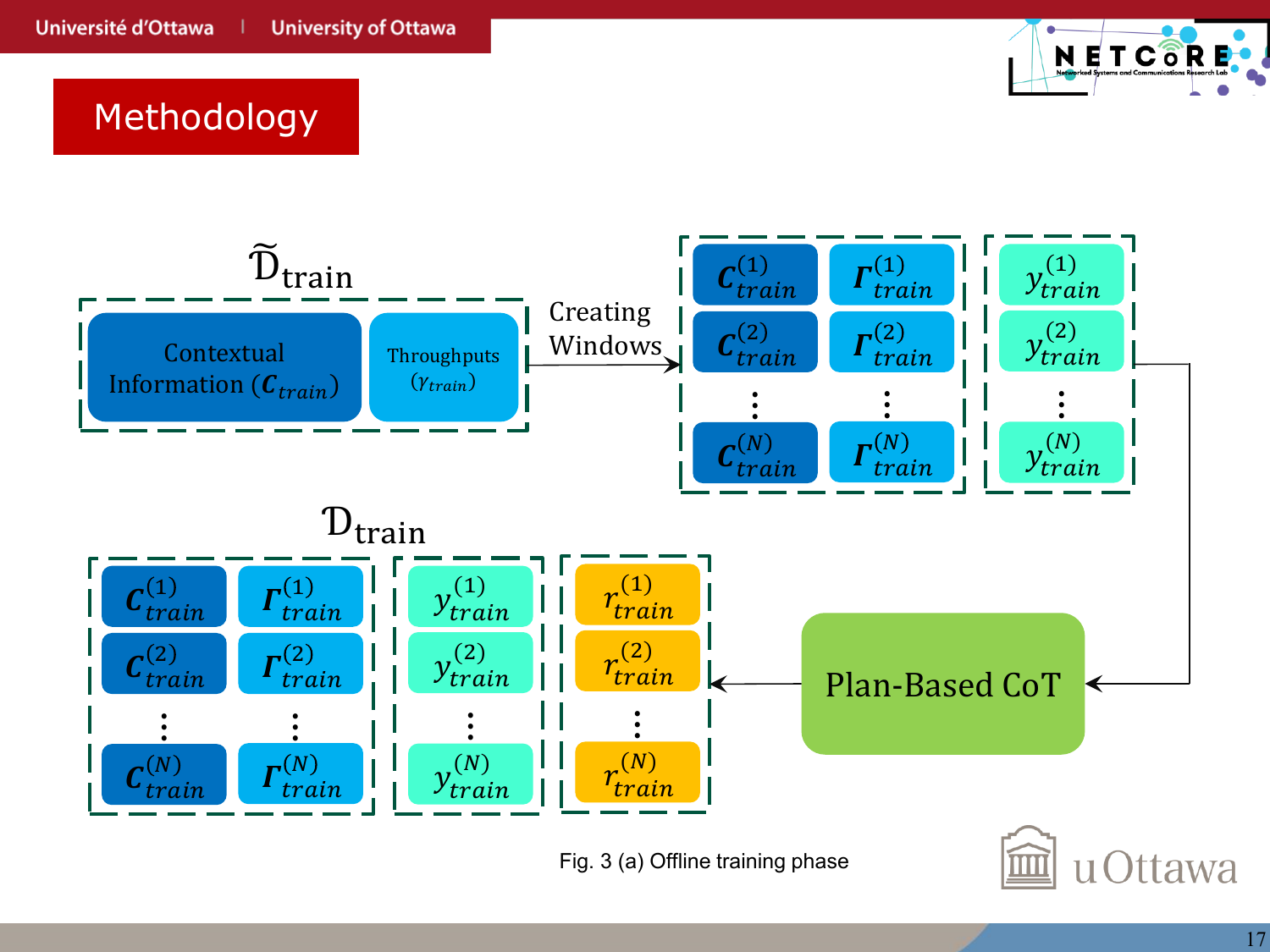}
    }
    \hfill
    \subfloat[\label{fig:myfig:b}]{
        \includegraphics[trim=16 80 19 106,clip,width=0.48\linewidth]{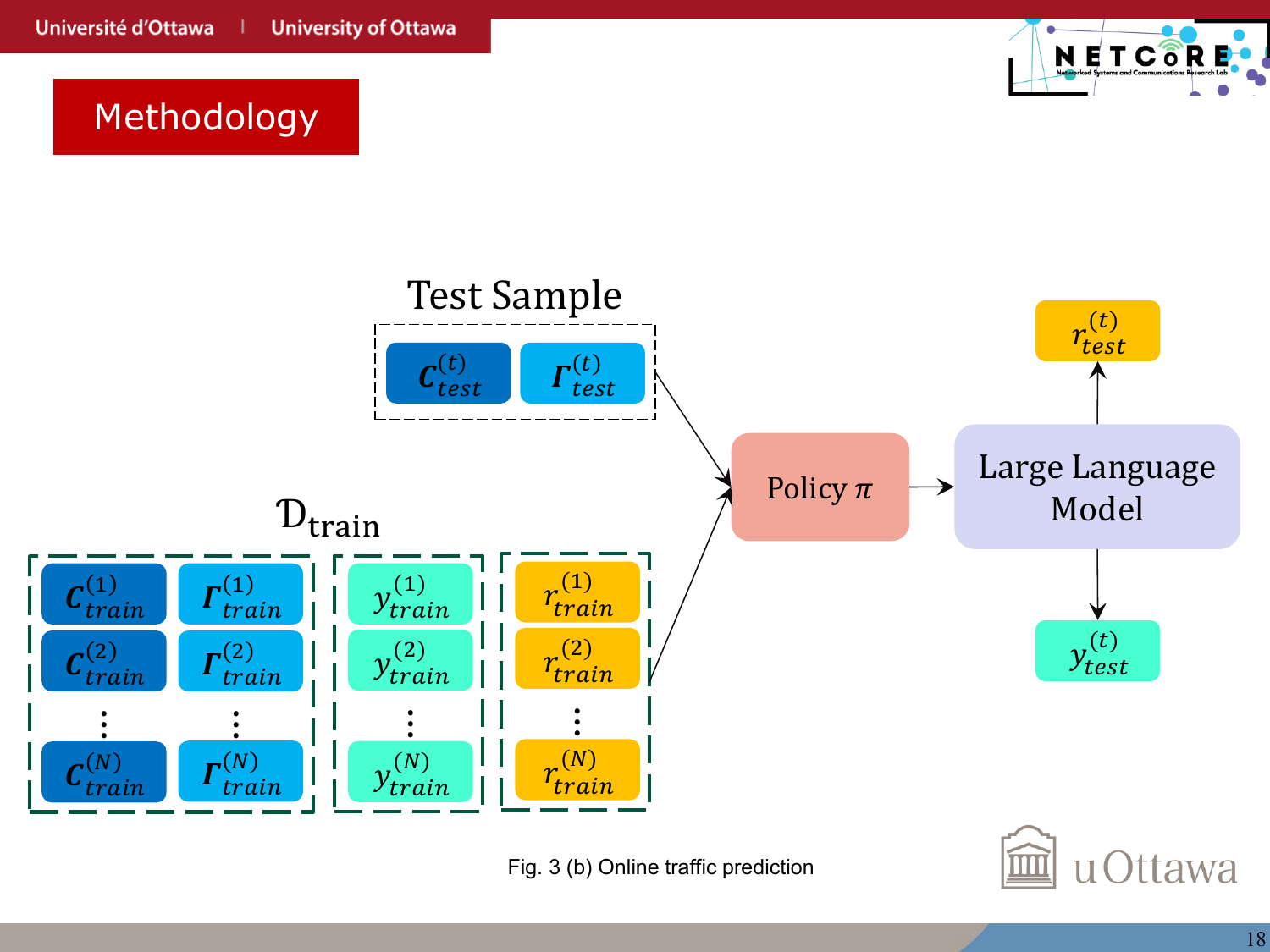}
    }

    \caption{Block diagram of the proposed few-shot \gls{CoT}-\gls{LLM} mobile traffic prediction framework. (a) offline prompt construction phase: traffic measurements are segmented into windows and fed through the \gls{PCoT} pipeline to generate structured training examples with \emph{rationales}. (b) Online traffic prediction phase: The selection policy $\pi$ retrieves the $M$ most relevant training examples, injects them into the \gls{LLM} prompt, and produces the predicted throughput.}
    \label{fig:block_diagram}
\end{figure*}

where $\mathbf{E}_{\text{train}}^{(n)}$ denotes the $n^{\text{th}}$ example in the processed training set $\mathcal{D}_{\text{train}} = \left\{\mathbf{E}_{\text{train}}^{(n)}\right\}_{n=1}^N$, where $n \in \left\{1,\ldots,N\right\}$ indexes the $N$ training examples. The specific methodology for generating the \emph{rationales} $r_{\text{train}}^{(n)}$ will be discussed in Section \ref{sec:methodology}. Subsequently, for $M$-shot \gls{CoT} prediction on a test sample $\mathbf{T}^{(t)}_{\text{test}}=\left[\bm{\Gamma}_{\text{test}}^{(t)}, \mathbf{C}_{\text{test}}^{(t)}\right]$ at time step $t$, the $M$ examples' indices are selected using a selection policy $\pi$ as:
\begin{equation}
    \mathcal{I}\left(\mathbf{T}_{\text{test}}^{(t)}\right) = \pi\left(\mathbf{T}_{\text{test}}^{(t)},\mathcal{D}_{\text{train}}\right), \quad \left|\mathcal{I}\left(\mathbf{T}_{\text{test}}^{(t)}\right)\right| = M.
\end{equation}
Subsequently, the selected examples' set for the test sample $\mathbf{T}_{\text{test}}^{(t)}$ can be defined as:
\begin{equation}
\label{eq:D_CoT}
    \mathcal{D}_{\text{\gls{CoT}}}\left(\mathbf{T}_{\text{test}}^{(t)}\right) = \left\{\mathbf{E}_{\text{train}}^{(m)}; m\in \mathcal{I}\left(\mathbf{T}_{\text{test}}^{(t)}\right)\right\}.
\end{equation}
Consequently, the predicted traffic and the \emph{rationale} using the \gls{LLM} under $M$-shot \gls{CoT} prompting can be written as:
\begin{equation}
\label{eq:inference}
    \left[\hat{y}_{\text{test}}^{(t)}, r_{\text{test}}^{(t)}\right] = f\left(\mathbf{T}_{\text{test}}^{(t)},\mathcal{D}_{\text{\gls{CoT}}}\left(\mathbf{T}_{\text{test}}^{(t)}\right)\vert \bm{\Theta}\right),
\end{equation}
where $\hat{y}_{\text{test}}^{(t)}$ and $r_{\text{test}}^{(t)}$ denote the predicted traffic and corresponding \emph{rationale}. \par 
The traffic prediction task aims to minimize the discrepancy between the predicted downlink throughput and the corresponding ground-truth throughput. For a fixed prompting configuration (policy $\pi$ and $M$ examples) we estimate the expected prediction error on the test set using the empirical average as:

\begin{equation}
	\label{eq:test_obj}
	\mathbb{E}\left[\mathcal{L}\left(\hat{y}_{\mathrm{test}},y_{\mathrm{test}}\right)\right]  =\frac{1}{T}\sum_{t=1}^{T} \mathcal{L}\!\left(\hat{y}^{(t)}_{\mathrm{test}},\, y^{(t)}_{\mathrm{test}}\right),
\end{equation}
where $T$ is the length of the test set and $\mathcal{L}$ is the error function.

\par In this work, we employ \gls{CoT}-enabled \glspl{LLM} to optimize the loss function, aiming to accurately predict traffic at the subsequent time step. Specifically, the \gls{CoT} mechanism enables the model to decompose complex temporal dependencies into intermediate reasoning steps, thereby capturing subtle variations in traffic patterns. The following section explains the proposed \gls{CoT}-\gls{LLM} traffic prediction algorithm.



\section{\gls{CoT}-Enabled \gls{LLM}-Based Mobile Traffic Prediction}
\label{sec:methodology}
In this section, we will explain the proposed \gls{CoT}-\gls{LLM} mobile traffic prediction algorithm. As illustrated in Fig. \ref{fig:block_diagram}, the proposed algorithm consists of two distinct phases: (a) an offline prompt construction phase, in which a novel framework is developed to generate the \emph{rationales} required for \gls{CoT}-\gls{LLM}-based traffic prediction, and (b) an online traffic prediction phase, in which the most relevant examples are selected using the policy $\pi$ and injected into the prompt.
Each of these phases will be described in detail in the following subsections.

\subsection{Phase 1: Offline Prompt Construction Phase}
This phase focuses on transforming the raw training dataset $\tilde{\mathcal{D}}_{\text{train}}$ into the structured examples $\mathbf{E}_{\text{train}}^{(n)}$, forming the processed training dataset $\mathcal{D}_{\text{train}}$. As described earlier in Section \ref{sec:problem}, the historical downlink throughput $\bm{\Gamma}$ and associated contextual feature sequences $\mathbf{C}$ are formed using (\ref{eq:past_throughput}) and (\ref{eq:past_info}), respectively. For a raw training dataset $\tilde{\mathcal{D}}_{\text{train}}$ containing $H$ seconds of traffic measurements, the number of training examples $N$ obtained with a window size $W$ and stride $S$ can be expressed as:
\begin{equation}
\label{eq:N}
    N = \left\lfloor\frac{H-W}{S}\right\rfloor + 1.
\end{equation}

After forming the historical windows, we generate the \emph{rationales} for the training data. These \emph{rationales} can have a substantial impact on the performance of \gls{CoT}-\gls{LLM} mobile traffic prediction, as we will observe. In particular, rather than only providing the final answers in the examples, we include intermediate steps and the reasoning behind each response. In this way, the model can better capture the underlying patterns and decision process required for accurate prediction. \par
\begin{figure*}
    \centering
    \includegraphics[trim= 8 16.3 8 48.5, clip, width=0.8\textwidth]{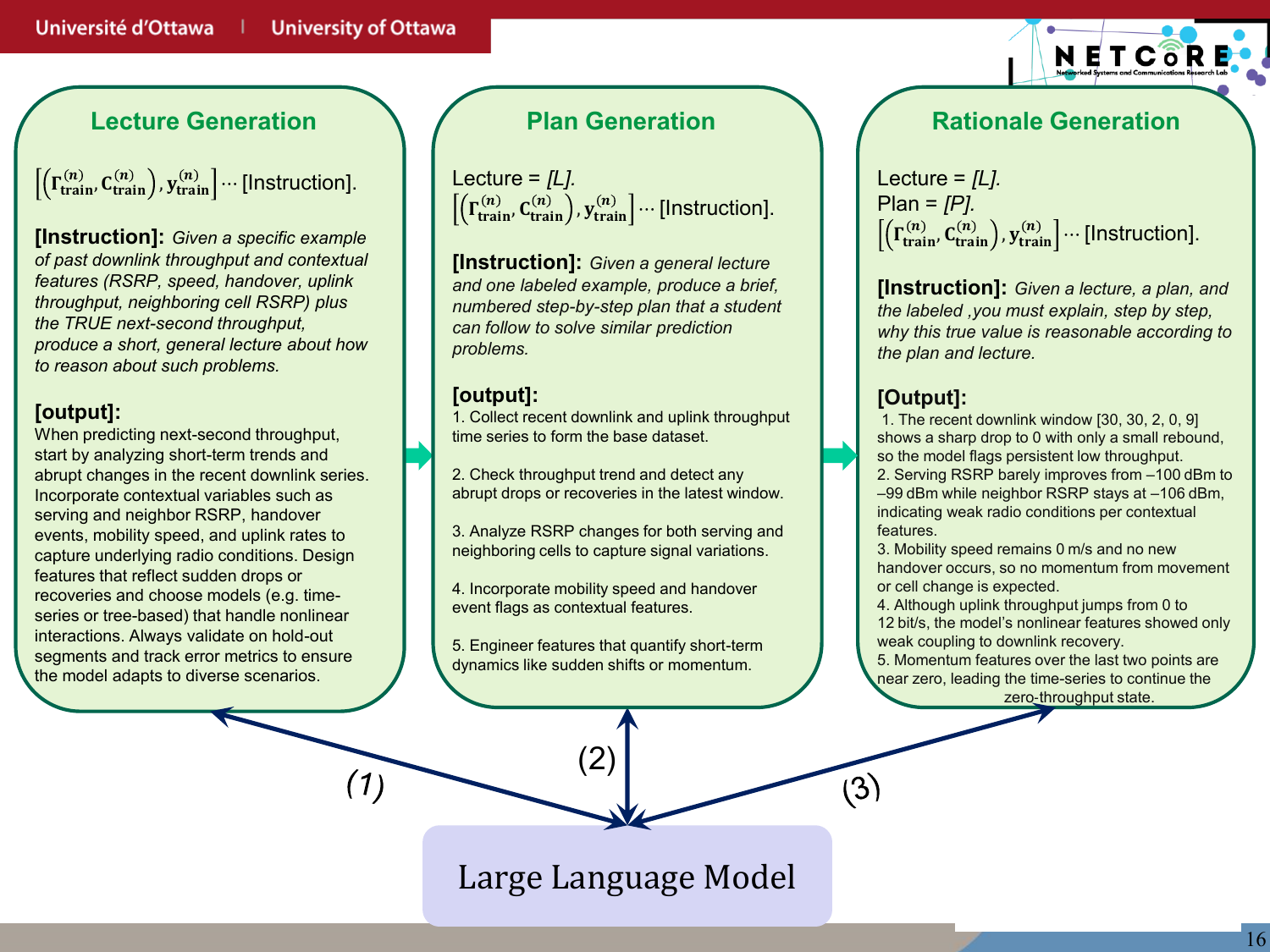}
    \caption{The block diagram of \gls{PCoT} for \emph{rationale} generation.}
    \label{fig:PCoT}
\end{figure*}

 
In this context, a straightforward approach for generating the \emph{rationales} is to rely on human expertise and manually craft several reasoning steps for each example. However, this solution can become impractical due to the large size of the training data, which makes the process time-consuming. Moreover, human knowledge may be insufficient to capture all aspects of the problem, potentially resulting in low-quality \emph{rationales}, and, consequently, performance degradation. Thus, we rely on \glspl{LLM} themselves for \emph{rationale} generation. In this case, we adopt a \gls{PCoT} strategy \cite{wang-etal-2023-plan,Huang2025ReasoningAI6G}, whereas the examination of other strategies is left as our future work. The \gls{PCoT} approach consists of three steps for \emph{rationale} generation, shown in Fig. \ref{fig:PCoT}, as:

\begin{enumerate}
    \item \textbf{Lecture Generation}: In this step, the model is prompted, using the \emph{instruction} $i_l$, to generate a general \emph{lecture} $l$ based on the past traffic throughput $\mathbf{\Gamma}_{\text{train}}^{(n)}$, the associated contextual information $\mathbf{C}_{\text{train}}^{(n)}$, and the corresponding ground-truth next-step traffic $y_{\text{train}}^{(n)}$ for the $n^{th}$ training example. This can be expressed as follows:
    \begin{equation}
    \label{eq:lecture}
        l = f\left(\left[\bm{\Gamma}_{\text{train}}^{(n)}, \mathbf{C}_{\text{train}}^{(n)}\right], y_{\text{train}}^{(n)}, i_l \vert \bm{\Theta}\right).
    \end{equation}
    \item \textbf{Plan Generation}: After acquiring the \emph{lecture}, the model is prompted to output a general \emph{plan} $p$, using the following:
    \begin{equation}
    \label{eq:plan}
        p = f\left(l, \left[\bm{\Gamma}_{\text{train}}^{(n)}, \mathbf{C}_{\text{train}}^{(n)}\right], y_{\text{train}}^{(n)},i_p \vert \bm{\Theta}\right),
    \end{equation}
    where $i_p$ denotes the \emph{instruction} for generating the \emph{plan}.
    \item \textbf{Rationale Generation}: Finally, using $l$ and $p$, the model is prompted to generate a \emph{rationale} as:
    \begin{equation}
    \label{eq:rationale}
        r_{\text{train}}^{(n)} = f\left(l, p, \left[\bm{\Gamma}_{\text{train}}^{(n)}, \mathbf{C}_{\text{train}}^{(n)}\right], y_{\text{train}}^{(n)},i_r \vert \bm{\Theta}\right).
    \end{equation}
    Here, $r_{\text{train}}^{(n)}$ represents the \emph{rationale} associated with the $n^{\text{th}}$ example, and $i_r$ is the \emph{instruction} included in the prompt to generate this \emph{rationale}. Using $r_{\text{train}}^{(n)}$, we then construct the complete $n^{\text{th}}$ training example $\mathbf{E}_{\text{train}}^{(n)}$ according to (\ref{eq:example}).
\end{enumerate}
Notably, the \emph{instructions} $i_l$, $i_p$, and $i_r$ are provided in Fig. \ref{fig:PCoT}. By repeating this process for all samples, we construct the processed training dataset $\mathcal{D}_{\text{train}}$ as:
\begin{equation}
    \mathcal{D}_{\text{train}} = \left\{\mathbf{E}_{\text{train}}\right\}_{n=1}^N.
\end{equation}
The summary of the offline prompt construction phase can be found in Algorithm \ref{alg:offline_phase}.
\begin{algorithm}[!t]
\caption{Offline Prompt Construction Phase}
\label{alg:offline_phase}
\SetAlgoLined

\KwIn{$\hat{\mathcal{D}}_{\text{train}}$, $H$, $W$, $S$, $f\left(.\vert\bm{\Theta}\right)$ $i_l$, $i_p$, and $i_r$.}
\KwOut{$\mathcal{D}_{\text{train}}$.}
\BlankLine
\BlankLine
Initialize $\mathcal{D}_{\text{temp}} \leftarrow \emptyset$\;
$n \leftarrow 0$\;
\BlankLine
\For{$t = W;~ t \le H-1;~ t \leftarrow t + S$}{
    $n \leftarrow n + 1$\;
    Construct $\bm{\Gamma}^{(t)}$ and $\mathbf{C}^{(t)}$ using (\ref{eq:past_throughput}) and (\ref{eq:past_info})\;
    $\bm{\Gamma}^{n}_{\text{train}} \leftarrow \bm{\Gamma}^{(t)}$\;
    $\mathbf{C}^{n}_{\text{train}} \leftarrow \mathbf{C}^{(t)}$\;
    $y^{n}_{\text{train}} \leftarrow \gamma^{(t+1)}$\;
    $\mathcal{D}_{\text{temp}} \leftarrow \mathcal{D}_{\text{temp}} \cup \{(\mathbf{C}^{n}_{\text{train}},\bm{\Gamma}^{n}_{\text{train}},y^{n}_{\text{train}})\}$\;
}
\BlankLine
$N \leftarrow n$\;
Initialize $\mathcal{D}_{\text{train}} \leftarrow \emptyset$\;
\BlankLine
\For{$n = 1;~ n \le N;~ n \leftarrow n + 1$}{
    Retrieve $(\mathbf{C}^{n}_{\text{train}},\bm{\Gamma}^{n}_{\text{train}},y^{n}_{\text{train}})$ from $\mathcal{D}_{\text{temp}}$\;
    Generate $l$ using (\ref{eq:lecture})\;
    Generate $p$ using (\ref{eq:plan})\;
    Generate $r^{n}_{\text{train}}$ using (\ref{eq:rationale})\;\
    Construct $\mathbf{E}^{n}_{\text{train}}$ using (\ref{eq:example})\;
    $\mathcal{D}_{\text{train}} \leftarrow \mathcal{D}_{\text{train}} \cup \{\mathbf{E}^{n}_{\text{train}}\}$\;
}
\BlankLine
\Return{$\mathcal{D}_{\text{train}}$}\;
\end{algorithm}
\subsection{Phase 2: Online Traffic Prediction}
This phase involves online traffic prediction, where using the policy $\pi$, we select the best-$M$ examples for $M$-shot \gls{CoT}-\gls{LLM} prediction. \par
In the first step, we explain the policy $\pi$, which follows a rule to retrieve examples that are most \emph{relevant} to the current test window. In particular, given the test input pair $\mathbf{T}_{\mathrm{test}}^{(t)} = \big[\bm{\Gamma}^{(t)}_{\mathrm{test}},\mathbf{C}^{(t)}_{\mathrm{test}}\big]$ at time step $t$, we score each candidate training example $\big[\bm{\Gamma}^{(n)}_{\mathrm{train}},\mathbf{C}^{(n)}_{\mathrm{train}}\big]$, $n\in\{1,\dots,N\}$, using a two-part distance that jointly captures the similarity of (i) the raw historical downlink throughput shape and (ii) its short-term dynamics (first-order increments). This design is motivated by the observation that, under an ICL-as-kernel-regression view, demonstrations that are \emph{more similar} to the test input tend to yield smaller prediction errors. \par
We first define the incremental (first-difference) throughput vector associated with a $W$-second historical window as:
\begin{equation}
\begin{split}
\label{eq:deltaGamma_online}
\Delta\bm{\Gamma}^{(t)} =
\Big[\gamma^{(t)}&-\gamma^{(t-1)},\; \gamma^{(t-1)}-\gamma^{(t-2)},\;\ldots \\ & \ldots, \;\gamma^{(t-W+2)}-\gamma^{(t-W+1)}\Big] \in \mathbb{R}^{W-1}.
\end{split}
\end{equation}

Then, for each training candidate $n$, we compute two Euclidean distances as follows:
\begin{align}
\label{eq:e1_online}
e_{1}(t,n) &= \left\|\bm{\Gamma}^{(t)}_{\mathrm{test}}-\bm{\Gamma}^{(n)}_{\mathrm{train}}\right\|_{2},\\
\label{eq:e2_online}
e_{2}(t,n) &= \left\|\Delta\bm{\Gamma}^{(t)}_{\mathrm{test}}-\Delta\bm{\Gamma}^{(n)}_{\mathrm{train}}\right\|_{2},
\end{align}
and combine them into a single score \cite{HanICL}:
\begin{equation}
\label{eq:effectiveness_score}
\mathcal{E}(t,n)= e_{1}(t,n)+e_{2}(t,n).
\end{equation}

Finally, the policy $\pi$ selects the indices of the $M$ most effective examples (smallest $\mathcal{E}(t,n)$):
\begin{equation}
\label{eq:pi_effectiveness}
\mathcal{I}\left(\mathbf{T}_{\mathrm{test}}^{(t)}\right)= \pi\!\left(\mathbf{T}_{\mathrm{test}}^{(t)}, \mathcal{D}_{\mathrm{train}} \right)
= \arg\min_{\substack{\mathcal{I}\subseteq\{1,\ldots,N\}\\|\mathcal{I}|=M}} \;\sum_{n\in\mathcal{I}} \mathcal{E}(t,n).
\end{equation}

The resulting $M$ examples are then assembled (together with their labels/rationales produced in Phase~1) to form $\mathcal{D}_{\mathrm{\gls{CoT}}}\left(\mathbf{T}_{\mathrm{test}}^{(t)}\right)$ using (\ref{eq:D_CoT}). Consequently, these components are injected into the prompt using (\ref{eq:inference}) for the \gls{CoT}-\gls{LLM} inference. The summary of the online phase is provided in Algorithm \ref{alg:online_phase}.
\begin{algorithm}[!t]
\caption{Online Prediction Phase}
\label{alg:online_phase}
\SetAlgoLined

\KwIn{$\mathcal{D}_{\text{train}}$, $\mathbf{T}_{\mathrm{test}}^{(t)}$, $M$, $\pi$, and $f\left(.\vert\bm{\Theta}\right)$.}
\KwOut{$\left[\hat{y}_{\mathrm{test}}^{(t)}, r^{(t)}_{\mathrm{test}}\right]$.}
\BlankLine
\BlankLine
Construct $\Delta\bm{\Gamma}_{\mathrm{test}}^{(t)}$ using (\ref{eq:deltaGamma_online})\;
\BlankLine
\For{$n = 1;~ n \le N;~ n \leftarrow n + 1$}{
    Retrieve $\bm{\Gamma}^{(n)}_{\text{train}}$ from $\mathcal{D}_{\text{temp}}$\;
    Construct $\Delta\bm{\Gamma}^{(n)}_{\mathrm{train}}$ using (\ref{eq:deltaGamma_online})\;
    set $e_1\left(t, n\right)$ using (\ref{eq:e1_online})\;
    set $e_2\left(t, n\right)$ using (\ref{eq:e2_online})\;
    Form $\mathcal{E}(t,n)$ using (\ref{eq:effectiveness_score})\;
}
\BlankLine
Select indices $\mathcal{I}\left(\mathbf{T}_{\mathrm{test}}^{(t)}\right)$ using (\ref{eq:pi_effectiveness})\;
Form $\mathcal{D}_{\mathrm{\gls{CoT}}}\left(\mathbf{T}_{\mathrm{test}}^{(t)}\right)$ using (\ref{eq:D_CoT})\;
Generate $\left[\hat{y}_{\mathrm{test}}^{(t)}, r^{(t)}_{\mathrm{test}}\right]$ using (\ref{eq:inference})\;
\BlankLine
\Return{$\left[\hat{y}_{\mathrm{test}}^{(t)}, r^{(t)}_{\mathrm{test}}\right]$}\;
\end{algorithm}


\section{NUMERICAL RESULTS}
\label{sec:results}
\begin{figure*}[!t]
    \centering

    \subfloat[Download, driving.\label{fig:vert_a}]{
        \includegraphics[width=0.65\textwidth]{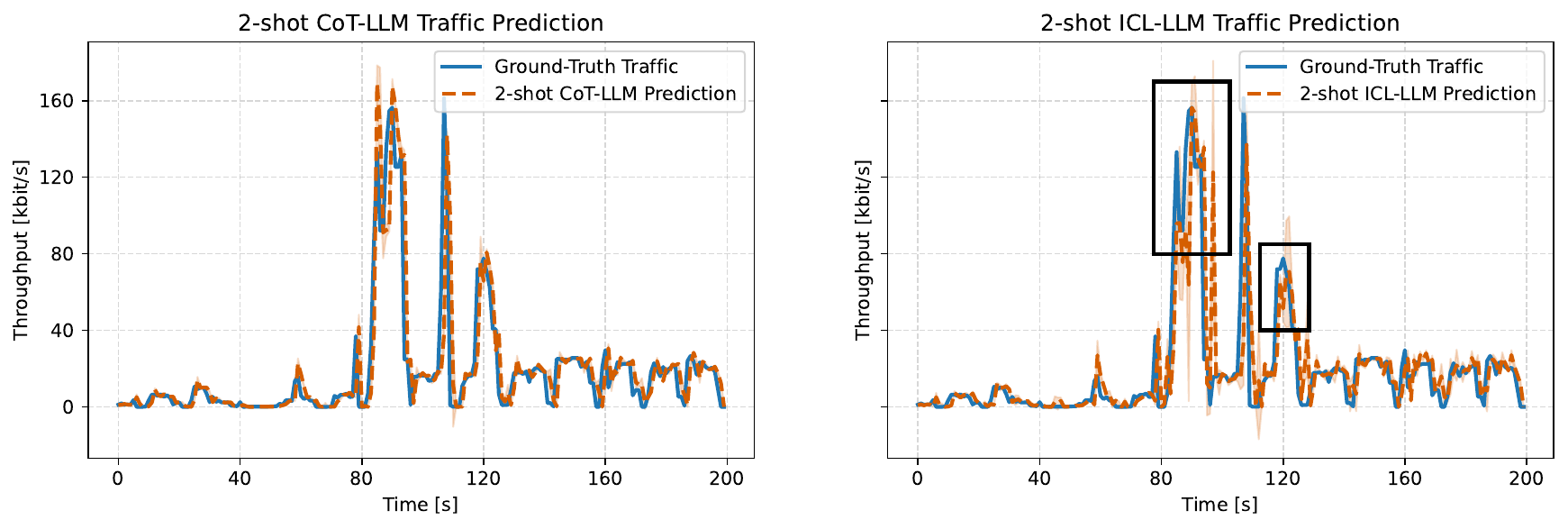}
    }\\[2mm]

    \subfloat[Watching Amazon Prime, driving.\label{fig:Amazon_traffic}]{
        \includegraphics[width=0.65\textwidth]{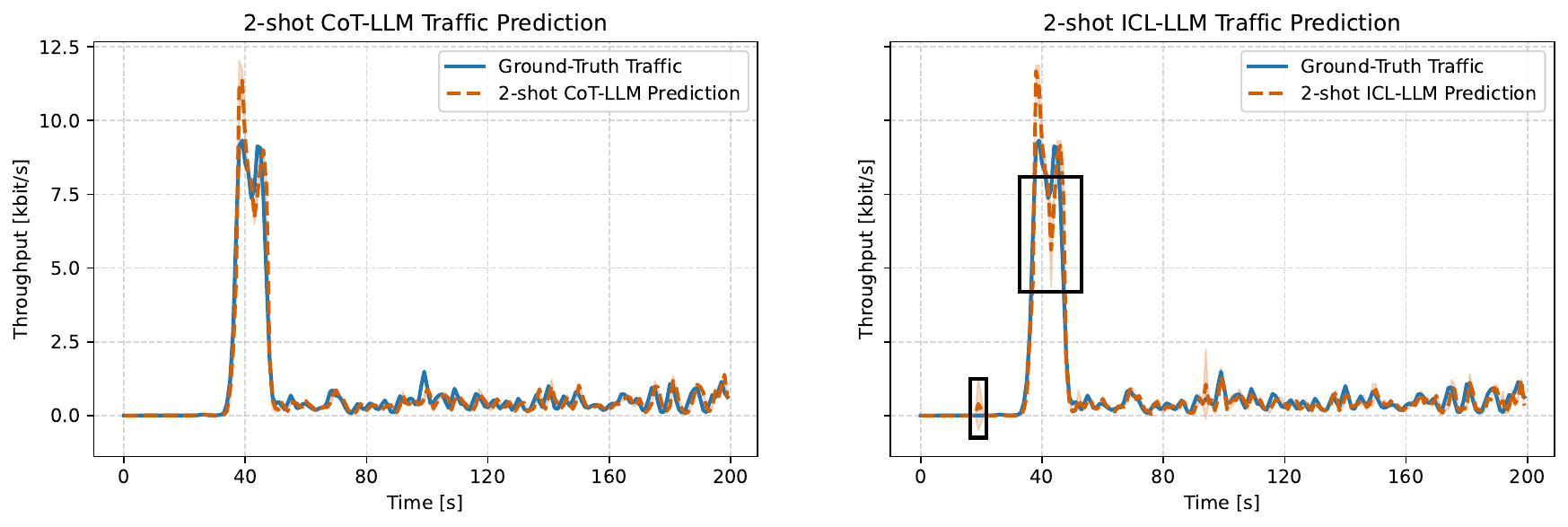}
    }\\[2mm]

    \subfloat[Download, static.\label{fig:vert_c}]{
        \includegraphics[width=0.65\textwidth]{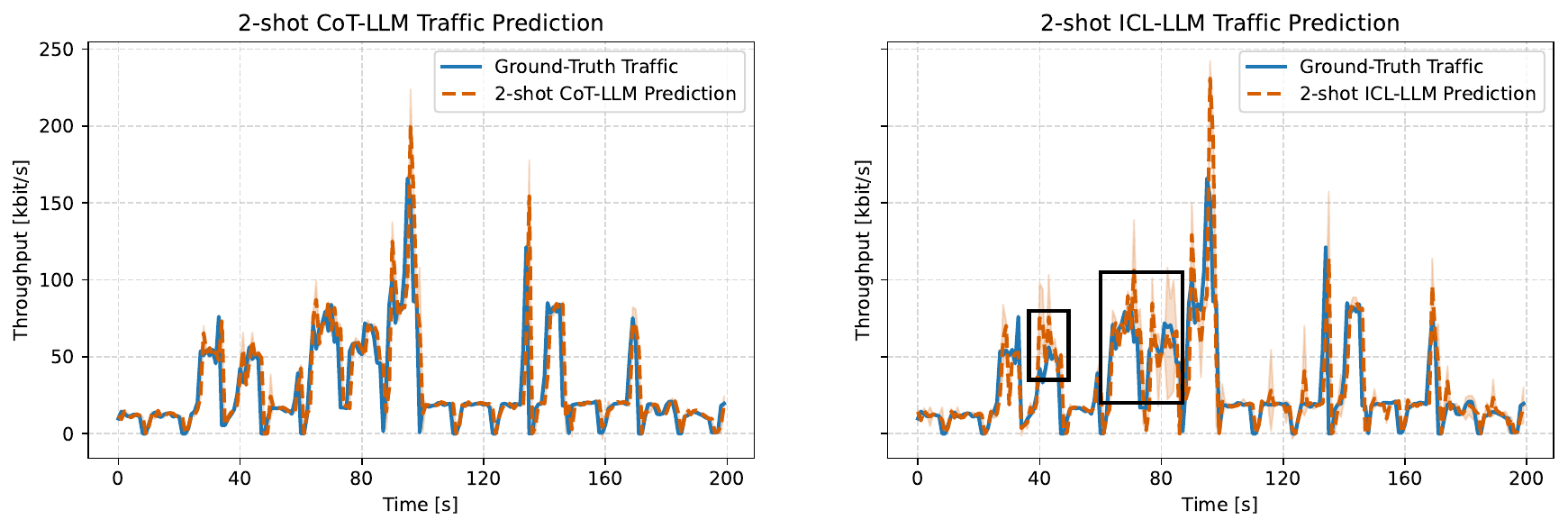}
    }

    \caption{Predicted traffic versus ground-truth traffic for 2-shot \gls{CoT}-\gls{LLM} and 2-shot \gls{ICL}-\gls{LLM}.}
    \label{fig:scenarios}
\end{figure*}
In this section, we present the numerical results obtained from our simulations. We investigate downlink mobile traffic prediction for real-world \gls{5G} user services using the measurement dataset in \cite{raca2020beyond}, which includes both static and driving scenarios across diverse applications, such as file downloading and video streaming (e.g., Amazon Prime), captured from operational networks. The dataset is publicly available and provides downlink throughput measurements together with various network-related contextual features, including channel metrics and neighboring-cell metrics. Specifically, it contains four categories of contextual information—throughput measurements, channel-related metrics, neighboring-cell metrics, and additional context indicators—resulting in a total of $25$ contextual features. For the downlink traffic prediction task, in addition to the historical downlink throughput, we focus on $K=5$ contextual features: uplink throughput, the \gls{RSRP} of the serving cell, the \gls{RSRP} of a neighboring cell, the network mode, and handover occurrence. This feature selection follows the feature-ranking analysis reported in \cite{Mei2022Realtime}, where importance scores were computed using the same dataset, and the five most impactful features were identified. \par
For these experiments, we use the o4-mini model \cite{openai2025o4mini} as the main \gls{LLM} and the same model is employed in both the offline and online phases. To improve the reliability of the results, all simulations are conducted over 5 independent runs, and the reported performance metrics are averaged across these runs. We split the raw dataset equally into training and testing subsets with no overlap between them, where the first $T=200$ seconds of each test trace are used for testing, and the remaining samples are reserved for training. The evaluation is conducted across multiple traffic scenarios, including downloading while driving, watching Amazon Prime while driving, and downloading in a static setting. Under this setup, the training data consist of 
$H=664$ seconds for the downloading-while-driving scenario, 
$H=1569$ seconds for the downloading-in-static scenario, and 
$H=588$ seconds for the Amazon Prime–while-driving scenario. Moreover, for data processing, we use a window size of \(W=5\) and a stride of \(S=1\) for all results reported in this section.
\subsection{Evaluation Metrics}
To evaluate the performance of various traffic prediction algorithms, we consider three main metrics as:
\begin{itemize}
    \item \textbf{\gls{MAE}: }This metric quantifies the average absolute deviation between the predicted and ground-truth values. It is computed as:
    \begin{equation}
        \text{\gls{MAE}} = \frac{1}{T}\sum\limits_{t=1}^T\left| \hat{y}^{(t)}-y^{(t)}\right|,
    \end{equation}
    where higher \gls{MAE} indicates lower prediction precision.

    \item \textbf{\gls{RMSE}: }This metric calculates the standard deviation of the predicted value, where similar to \gls{MAE}, higher values show weaker predictions. This metric can be expressed as:
    \begin{equation}
        \text{\gls{RMSE}} = \sqrt{\frac{1}{T}\sum\limits_{t=1}^T\left( \hat{y}^{(t)}-y^{(t)}\right)^2}\ .
    \end{equation}
    \item \textbf{$\mathbf{R^2}$-score: }This metric measures the proportion of the variance in the ground-truth values that is explained by the predictions, indicating how well the model fits the data. It can be calculated as:
    \begin{equation}
R^2\mathrm{-score} = 1 - \frac{\sum\limits_{t=1}^{T}\left(y^{\left(t\right)} - \hat{y}^{\left(t\right)}\right)^2}{\sum\limits_{t=1}^{T}\left(y^{\left(t\right)} - \bar{y}\right)^2}, \quad \bar{y} = \frac{1}{T}\sum\limits_{t=1}^Ty^{\left(t\right)}.
    \end{equation}
The $R^2$-score ranges between $-\infty$ to $1$, where higher values are an indication of a better prediction. Notably, an $R^2$-score of $0$ indicates that the model performs no better than simply predicting the mean of the ground-truth values.
\end{itemize}
\subsection{\gls{CoT}-\gls{LLM} Mobile Traffic Prediction}
In the first step, we evaluate the performance of the proposed \gls{CoT}-\gls{LLM} mobile traffic prediction approach against existing benchmarks across multiple scenarios. Fig. \ref{fig:scenarios} illustrates the predicted traffic alongside the ground-truth traffic for three cases—namely, downloading while driving, watching Amazon Prime while driving, and downloading in a static setting—where a 2-shot \gls{CoT}-\gls{LLM} is compared with the 2-shot \gls{ICL}-\gls{LLM} baseline in \cite{HanICL}, where the authors consider plain \gls{ICL} using a similar example selection policy for traffic prediction. At this stage, we consider a 2-shot setup (two in-prompt examples) since it achieved an acceptable performance in \cite{HanICL}. We discuss the impact of number of examples $M$ in the later subsections. It can be seen that the proposed 2-shot \gls{CoT}-\gls{LLM} achieve a better prediction compared to the 2-shot \gls{ICL}-\gls{LLM}, where the higher error parts are highlighted for \gls{ICL}-\gls{LLM} in Fig. \ref{fig:scenarios}. \par 

\begin{table*}[t!]
    \centering
    \caption{Summary of performance on test data across scenarios (mean $\pm$ std over 5 runs). 
    }
    \label{tab:perf_three_scenarios}
    \resizebox{\textwidth}{!}{%
    \begin{tabular}{c|ccc|ccc|ccc}

    \multirow{2}{*}{\textbf{Method}}
    & \multicolumn{3}{c|}{\textbf{Download (Driving)}}
    & \multicolumn{3}{c|}{\textbf{Amazon Prime (Driving)}}
    & \multicolumn{3}{c}{\textbf{Download (Static)}} \\
    \cline{2-10}

    &  \textbf{MAE }$\downarrow$  & \textbf{RMSE }$\downarrow$ & $\mathbf{R^2}\uparrow$
    &  \textbf{MAE }$\downarrow$  & \textbf{RMSE }$\downarrow$ & $\mathbf{R^2}\uparrow$
    &  \textbf{MAE }$\downarrow$  & \textbf{RMSE }$\downarrow$ & $\mathbf{R^2}\uparrow$ \\
    \hline

    \rowcolor{myyellow}
    2-shot \gls{CoT}-\gls{LLM} (ours)
    & $\mathbf{8.039 \pm 0.257}$ & $\mathbf{18.552 \pm 0.377}$ & $\mathbf{0.639 \pm 0.015}$
    & $0.230 \pm 0.008$ & $\mathbf{0.447 \pm 0.025}$ & $\mathbf{0.936 \pm 0.007}$
    & $\mathbf{9.799 \pm 0.303}$ & $\mathbf{21.237 \pm 0.898}$ & $\mathbf{0.458 \pm 0.046}$ \\
    \hline
    2-shot \gls{ICL}-\gls{LLM} \cite{HanICL}
    & $9.235 \pm 0.216$ & $21.341 \pm 0.17$ & $0.522 \pm 0.053$
    & $\mathbf{0.222 \pm 0.014}$ & $0.489 \pm 0.029$ & $0.924 \pm 0.009$
    & $11.171 \pm 0.334$ & $22.324 \pm 1.458$ & $0.399 \pm 0.078$ \\
    \hline
    Zero-shot \gls{CoT}-\gls{LLM}
    & $8.891 \pm 0.380$ & $22.058 \pm 2.940$ & $0.483 \pm 0.140$
    & $0.254 \pm 0.014$ & $0.481 \pm 0.038$ & $0.926 \pm 0.012$
    & $12.770 \pm 0.454$ & $23.760 \pm 1.194$ & $0.321 \pm 0.067$ \\
    \hline
    Zero-shot \gls{ICL}-\gls{LLM} \cite{HanICL}
    & $9.424 \pm 0.533$ & $23.214 \pm 4.193$ & $0.421 \pm 0.220$
    & $0.269 \pm 0.022$ & $0.528 \pm 0.064$ & $0.910 \pm 0.021$
    & $13.779 \pm 0.543$ & $25.275 \pm 1.303$ & $0.232 \pm 0.077$ \\
    \hline
    \gls{SMA}
    & $12.385 \pm 0.000$ & $24.745 \pm 0.000$ & $0.359 \pm 0.000$
    & $0.419 \pm 0.000$ & $1.098 \pm 0.000$ & $0.615 \pm 0.000$
    & $16.412 \pm 0.000$ & $25.168 \pm 0.000$ & $0.240 \pm 0.000$ \\
    \hline
    \gls{WMA} \cite{10.1007/s10922-016-9392-x}
    & $10.584 \pm 0.000$ & $22.041 \pm 0.000$ & $0.491 \pm 0.000$
    & $0.365 \pm 0.000$ & $0.915 \pm 0.000$ & $0.732 \pm 0.000$
    & $14.051 \pm 0.000$ & $22.927 \pm 0.000$ & $0.369 \pm 0.000$ \\
    \hline
    \gls{ARIMA} \cite{TianL21-1}
    & $10.757 \pm 0.000$ & $22.385 \pm 0.580$ & $0.466 \pm 0.000$
    & $0.288 \pm 0.000$ & $0.665 \pm 0.000$ & $0.859 \pm 0.000$
    & $14.854 \pm 0.000$ & $25.034 \pm 0.000$ & $0.248 \pm 0.000$ \\
    \hline
    Kalman Filter
    & $10.400 \pm 0.000$ & $21.488 \pm 0.000$ & $0.516 \pm 0.000$
    & $0.355 \pm 0.000$ & $0.896 \pm 0.000$ & $0.744 \pm 0.000$
    & $13.775 \pm 0.000$ & $22.259 \pm 0.000$ & $0.405 \pm 0.000$ \\
    \end{tabular}%
    }
\end{table*}


To further assess performance, we report the \gls{MAE}, \gls{RMSE}, and $R^2$-score of the 2-shot \gls{CoT}-\gls{LLM} and 2-shot \gls{ICL}-\gls{LLM} \cite{HanICL} in Table \ref{tab:perf_three_scenarios}. Furthermore, in this table, we consider more benchmarks: 

\begin{enumerate}
    \item \textbf{zero-shot \gls{CoT}-\gls{LLM}:} The \gls{CoT} prompting without injecting examples, where the model is explicitly asked to \emph{think step-by-step} before outputting the traffic. 
    \item \textbf{zero-shot \gls{ICL}-\gls{LLM}:} The plain \gls{ICL} without the in-context examples discussed in \cite{HanICL}.
    \item \textbf{\gls{SMA}: } The average of $W = 5$ seconds of previous traffic throughput.
    \item \textbf{\gls{WMA} \cite{10.1007/s10922-016-9392-x}:} The weighted average with incremental weights of the $W = 5$ seconds of past traffic.
    \item \textbf{\gls{ARIMA}:} The auto regressive-based traffic prediction approach discussed in \cite{TianL21-1}.
    \item \textbf{Kalman Filter:} The Kalman filter using a local level model with recursive state updates.
\end{enumerate}

\par 
Overall, the proposed 2-shot \gls{CoT}-\gls{LLM} achieves the best performance across considered scenarios. For instance, over downloading while driving setting, the proposed 2-shot \gls{CoT}-\gls{LLM} algorithm attains an average \gls{MAE} of $8.039$, \gls{RMSE} of $18.552$, and $R^2$-score of $0.639$, whereas the next-best algorithm, 2-shot \gls{ICL}-\gls{LLM}, achieves an average $9.235$ \gls{MAE}, $21.341$ \gls{RMSE}, and $0.522$ $R^2$-score. This means that using \gls{CoT}-\gls{LLM} can boost the \gls{MAE}, \gls{RMSE}, and $R^2$-score by $14.88\%$, $15.03\%$, and $22.41\%$, respectively. While a similar trend can be observed in the other settings, it is worth noting that only in watching Amazon Prime while driving the 2-shot \gls{ICL}-\gls{LLM} \cite{HanICL} achieves a slightly lower \gls{MAE} (by $3.60\%$); however, the proposed 2-shot \gls{CoT}-\gls{LLM} still improves the \gls{RMSE} and $R^2$-score by $9.4\%$ and $1.3\%$, respectively. This can be attributed to the fact that this scenario, illustrated in Fig. \ref{fig:Amazon_traffic}, is easier to predict; consequently, the performance of \gls{ICL}-\gls{LLM} and \gls{CoT}-\gls{LLM} is closer to to each other.
\par 
This table also also highlights the benefit of few-shot learning compared to zero-shot learning. Particularly, for instance in downloading while driving scenario, the 2-shot \gls{CoT}-\gls{LLM} improves upon the zero-shot \gls{CoT}-\gls{LLM} by $10.60\%$, $18.90\%$, and $32.30\%$ in terms of \gls{MAE}, \gls{RMSE}, and $R^2$-score, respectively. A similar improvement trend is also observed when comparing 2-shot \gls{ICL}-\gls{LLM} and zero-shot \gls{ICL}-\gls{LLM}. \par

Compared to classical benchmarks, the performance gap becomes even more evident. For example, in downloading while static setup, the 2-shot \gls{CoT}-\gls{LLM} improves the \gls{MAE}, \gls{RMSE}, and $R^2$-score by $38.26\%$, $5.82\%$, and $17.89\%$, respectively, compared to \gls{WMA} \cite{10.1007/s10922-016-9392-x}. Similarly, for the same setup, the 2-shot \gls{CoT}-\gls{LLM} enhances the Kalman  filter performance by $35.34\%$, $15.58\%$, and $7.41\%$ on \gls{MAE}, \gls{RMSE}, and $R^2$-score, respectively. \par
It is important to note that the achieved ranges of \gls{MAE}, \gls{RMSE}, and \(R^2\)-score vary across different scenarios. This is mainly due to differences in the downlink throughput range and the level of traffic fluctuation in each setting, which directly affect the difficulty of the prediction task. For instance, in the Amazon Prime while driving scenario, all benchmarks attain relatively high \(R^2\)-scores and low \gls{RMSE} and \gls{MAE}, indicating a more predictable traffic pattern. In contrast, for downloading in a static scenario, the \(R^2\)-score drops to nearly half for all methods, which suggests a highly fluctuating traffic trend and a more challenging prediction environment.


\subsection{Ablation Study} 
In this subsection, we conduct an ablation study to systematically evaluate the contribution of each component in the proposed 2-shot \gls{CoT}-\gls{LLM} framework for mobile traffic prediction. Without loss of generality, the downloading-while-driving scenario is used as a representative example, as similar performance trends are observed across other scenarios. Specifically, we examine the impact of the following elements within the selection policy $\pi$ and the use of \emph{rationales}:
\begin{enumerate}
	\item The selection criterion $e_1$ in (\ref{eq:e1_online}).
	\item The selection criterion $e_2$ in (\ref{eq:e2_online}).
	\item The inclusion of \emph{rationales} in the prompt.
\end{enumerate}
\par
\begin{figure*}[t!]
	\centering
	
	\subfloat[\label{fig:ablation_rationale}]{
		\includegraphics[width=0.41\linewidth]{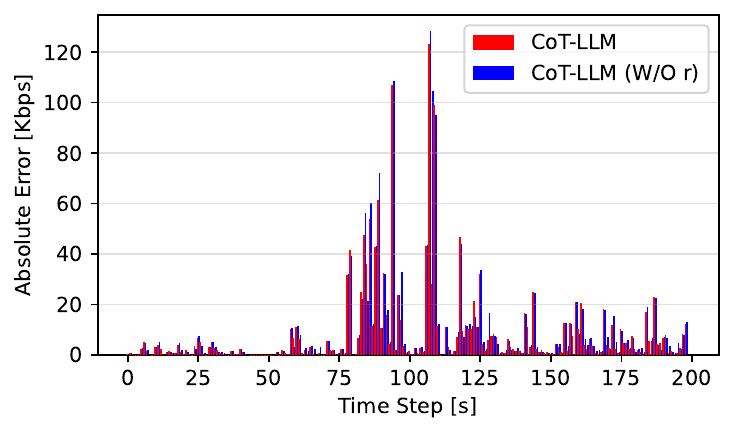}
	}
	\hfill
	\subfloat[\label{fig:ablation_selection}]{
		\includegraphics[width=0.41\linewidth]{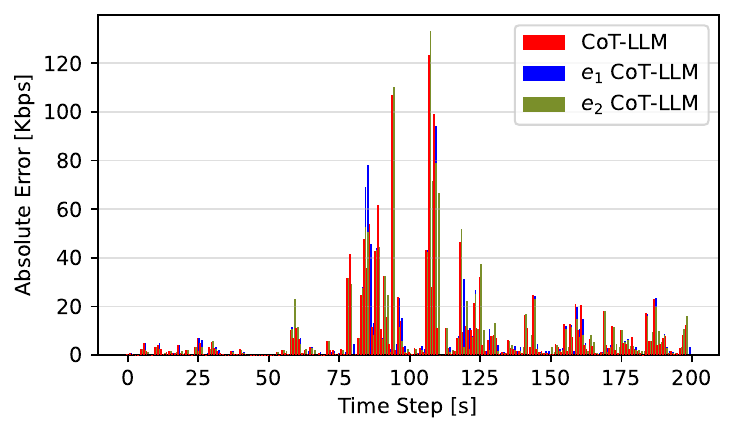}
	}
	
	\caption{The absolute error for (a) 2-shot \gls{CoT}-\gls{LLM} without \emph{rationales} and (b) various selection policies.}
	\label{fig:ablation}
\end{figure*}

 Figure~\ref{fig:ablation} illustrates the absolute prediction error for all considered variants. In particular, Fig.~\ref{fig:ablation_rationale} compares the proposed 2-shot \gls{CoT}-\gls{LLM} against a baseline in which the \emph{rationales} are omitted from the prompt, i.e.,
\begin{equation}
	\left[\hat{y}_{\text{test}}^{(t)}, r_{\text{test}}^{(t)}\right] = f\left(\mathbf{T}_{\text{test}}^{(t)}, \left[ \bm{\Gamma}, \mathbf{C}, y\right]\vert \bm{\Theta}\right).
\end{equation}
\par 
Similarly, Fig.~\ref{fig:ablation_selection} reports the absolute error when \emph{rationales} are retained, but either $e_1$ or $e_2$ is removed from the selection policy $\pi$. To quantitatively assess the contribution of each component, the corresponding \gls{MAE}, \gls{RMSE}, and $R^2$-score for all ablation variants are summarized in Table~\ref{tab:ablation}. \par
The results indicate that all components positively contribute to the overall prediction performance. For example, excluding the \emph{rationales} from the prompt leads to performance degradations of $-7.04\%$, $-17.34\%$, and $-29.09\%$ in terms of \gls{MAE}, \gls{RMSE}, and $R^2$-score, respectively, compared to the full 2-shot \gls{CoT}-\gls{LLM} configuration. Likewise, removing criterion $e_1$ from the selection policy $\pi$ results in performance drops of $-0.30\%$ in \gls{MAE}, $-3.35\%$ in \gls{RMSE}, and $-3.09\%$ in $R^2$-score. \par
Notably, the inclusion of \emph{rationales} yields the most significant performance gains. While modifications to the selection policy $\pi$ introduce measurable degradation, their impact is substantially smaller than that caused by removing the \emph{rationales}. This observation underscores the critical role of structured reasoning information in enhancing the effectiveness of \gls{CoT}-\gls{LLM} based prompting for mobile traffic prediction.

 \begin{table}[t!]
	\centering
	\caption{An ablation study on the impact of each component compared to the proposed 2-shot \gls{CoT}-\gls{LLM}.
    }
	\label{tab:ablation}
	\resizebox{\linewidth}{!}{%
		\begin{tabular}{c|ccc|ccc}
			
			\multirow{2}{*}{\textbf{Method}}
			& \multicolumn{3}{c|}{\textbf{Obtained}}
			& \multicolumn{3}{c}{\textbf{Average Change}} \\
			\cline{2-7}
			
			& \textbf{MAE }$\downarrow$ & \textbf{RMSE }$\downarrow$ & $\mathbf{R^2}\uparrow$
			& \textbf{MAE } & \textbf{RMSE } & $\mathbf{R^2}$ \\
			\hline

            \gls{CoT}-\gls{LLM}
			& $8.039 \pm 0.257$ & $18.552 \pm 0.377$ & $0.639 \pm 0.015$
			& $-$ & $-$ & $-$ \\
			\hline
            
			\gls{CoT}-\gls{LLM} (W/O $r$)
			& $8.605 \pm 0.605$ & $21.769 \pm 3.220$ & $0.495 \pm 0.149$
			& $\textcolor{red}{\mathbf{-7.04\%}}$ & $\textcolor{red}{\mathbf{-17.34\%}}$ & $\textcolor{red}{\mathbf{-29.09\%}}$ \\
			\hline
			
			$e_1$ \gls{CoT}-\gls{LLM}
			& $8.063 \pm 0.189$ & $19.174 \pm 0.245$ & $0.615 \pm 0.010$
			& $\textcolor{red}{\mathbf{-0.30\%}}$ & $\textcolor{red}{\mathbf{-3.35\%}}$ & $\textcolor{red}{\mathbf{-3.09\%}}$ \\
			\hline
			
			$e_2$ \gls{CoT}-\gls{LLM}
			& $8.327 \pm 0.202$ & $19.865 \pm 0.340$ & $0.601 \pm 0.017$
			& $\textcolor{red}{\mathbf{-3.58\%}}$ & $\textcolor{red}{\mathbf{-7.08\%}}$ & $\textcolor{red}{\mathbf{-6.32\%}}$ \\
			
		\end{tabular}%
	}
\end{table}

\subsection{Number of Examples}
\begin{figure}[t!]
	\centering
	\includegraphics[width=\linewidth]{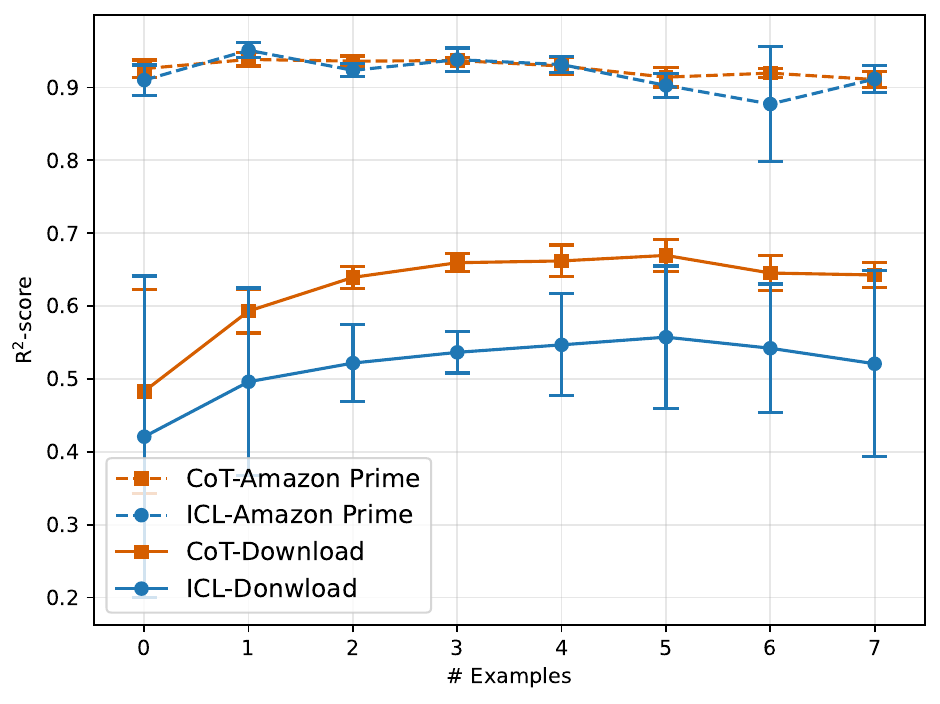}
	\caption{$R^2$-score versus number of examples $M$.}
	\label{fig:examples}
\end{figure}

This subsection investigates the impact of the number of in-context examples on the performance of \gls{CoT}-\gls{LLM} and \gls{ICL}-\gls{LLM}–based traffic prediction. The corresponding simulation results are presented in Fig.~\ref{fig:examples} and Table~\ref{tab:examples}. The results demonstrate that increasing the number of examples does not necessarily lead to consistent performance improvements for either approach. In fact, an excessive number of examples can degrade prediction accuracy.

For the downloading-while-driving scenario, performance improves as the number of examples increases up to $M=5$, beyond which further increases lead to noticeable degradation. By optimally selecting the number of examples, the prediction performance reaches \gls{MAE} = $7.687$, \gls{RMSE} = $17.755$, and an $R^2$-score of $0.670$. This corresponds to improvements of $4.58\%$, $5.70\%$, and $4.85\%$ in \gls{MAE}, \gls{RMSE}, and $R^2$-score, respectively, compared to the 2-shot \gls{CoT}
-\gls{LLM} baseline. In contrast, for the watching Amazon Prime while driving scenario, reducing the number of examples to $M=1$ yields superior overall performance. While this reduction leads to a $3.47\%$ degradation in \gls{MAE}, it improves the \gls{RMSE} and $R^2$-score by $2.05\%$ and $0.20\%$, respectively, as compared to 2-shot \gls{CoT}-\gls{LLM}.
\par

When compared to the \gls{ICL}-\gls{LLM} baseline, the proposed 5-shot \gls{CoT}-\gls{LLM} achieves a substantial performance gain in the downloading-while-driving scenario, outperforming \gls{ICL}-\gls{LLM} by $20.27\%$ in terms of the $R^2$-score. However, after optimizing the number of examples for both approaches, \gls{ICL}-\gls{LLM} exhibits a marginal $1.39\%$ improvement over \gls{CoT}-\gls{LLM} in watching Amazon Prime and driving setting. In particular, when comparing the \gls{PE} (which measures the irregularity of a time series) \cite{Bandt2002PermutationEntropy} of these two settings, we observe that the downloading while driving traffic exhibits a higher normalized \gls{PE} than the watching Amazon Prime while driving traffic, with $H_{\mathrm{PE}}^{\mathrm{norm}}=0.780$ and $H_{\mathrm{PE}}^{\mathrm{norm}}=0.701$, respectively\footnote{The normalized \gls{PE} value lies in $[0,1]$, where larger values indicate a higher diversity of local ordinal patterns and, therefore, a less locally predictable (more irregular) time series.}. This indicates a larger diversity of local ordinal patterns and, consequently, higher short-term irregularity in the downloading while driving setting. Such increased temporal complexity suggests that this scenario is intrinsically more challenging to predict, which is consistent with the larger performance gain achieved by the proposed \gls{CoT}-\gls{LLM} in this case. Conversely, the lower \gls{PE} of watching Amazon Prime while driving implies a relatively more regular structure, for which standard \gls{ICL}-\gls{LLM} prompting remains competitive and can yield a slight advantage after per-scenario tuning. Notably, the overall performance of the two methods becomes very close for the later setting, with differences remaining within the confidence intervals of both algorithms. \par
\begin{table*}[t!]
	\centering
	\caption{Summary of performance on test data with different number of examples (mean $\pm$ std over 5 runs).}
	\label{tab:examples}
	\resizebox{\textwidth}{!}{%
		\begin{tabular}{c|l|cccccccc}
			\hline
			
			\multicolumn{10}{c}{\textbf{Download (Driving)}}\\
			\hline
			\textbf{Method} & \textbf{Metric}
			& \textbf{$M=0$} & \textbf{$M=1$} & \textbf{$M=2$} & \textbf{$M=3$}
			& \textbf{$M=4$} & \textbf{$M=5$} & \textbf{$M=6$} & \textbf{$M=7$} \\
			\hline
			
			\multirow{3}{*}{$M$-shot \gls{CoT}-\gls{LLM}}
			& \textbf{MAE }$\downarrow$  & $8.891 \pm 0.380$ & $8.310 \pm 0.161$ & $8.039 \pm 0.257$ & $7.786 \pm 0.089$ & $7.853 \pm 0.315$ & $\mathbf{7.687 \pm 0.136}$ & $7.816 \pm 0.274$ & $7.748 \pm 0.092$ \\
			& \textbf{RMSE }$\downarrow$ & $22.058 \pm 2.940$ & $19.697 \pm 0.728$ & $18.551 \pm 0.377$ & $18.027 \pm 0.324$ & $17.957 \pm 0.578$ & $\mathbf{17.755 \pm 0.579}$ & $18.392 \pm 0.633$ & $18.465 \pm  0.437$ \\
			& $\mathbf{R^2}\uparrow$   & $0.483 \pm 0.140$ & $0.593 \pm 0.030$ & $0.639 \pm 0.015$ & $0.660 \pm 0.012$ & $0.662 \pm 0.022$ & $\mathbf{0.670 \pm 0.022}$ & $0.645 \pm 0.024$ & $0.643 \pm0.017$ \\
			\cline{1-10}
			
			\multirow{3}{*}{$M$-shot \gls{ICL}-\gls{LLM} \cite{HanICL}}
			& \textbf{MAE }$\downarrow$  & $9.424 \pm 0.533$ & $9.326 \pm 0.707$ & $9.235 \pm 0.216$ & $9.243 \pm 0.251$ & $9.168 \pm 0.477$ & $9.156 \pm 0.818$ & $9.228 \pm 0.575$ & $9.049 \pm 0.848$ \\
			& \textbf{RMSE }$\downarrow$ & $23.214 \pm 4.193$ & $21.782 \pm 2.854$ & $21.341 \pm 1.169$ & $21.027 \pm 0.638$ & $20.750 \pm 1.614$ & $20.461 \pm 2.214$ & $20.819 \pm 2.133$ & $21.232 \pm 2.879$ \\
			& $\mathbf{R^2}\uparrow$    & $0.421 \pm 0.220$ & $0.496 \pm 0.129$ & $0.522 \pm 0.053$ & $0.537 \pm 0.028$ & $0.547 \pm 0.070$ & $0.557 \pm 0.098$ & $0.542 \pm 0.088$ & $0.521 \pm 0.128$ \\
			\hline
			
			\multicolumn{10}{c}{\textbf{Amazon Prime (Driving)}}\\
			\hline
			\textbf{Method} & \textbf{Metric}
			& \textbf{$M=0$} & \textbf{$M=1$} & \textbf{$M=2$} & \textbf{$M=3$}
			& \textbf{$M=4$} & \textbf{$M=5$} & \textbf{$M=6$} & \textbf{$M=7$} \\
			\hline
			
			\multirow{3}{*}{$M$-shot \gls{CoT}-\gls{LLM}}
			& \textbf{MAE }$\downarrow$  & $0.254 \pm 0.014$ & $0.238 \pm 0.010$ & $0.230 \pm 0.008$ & $0.230 \pm 0.007$ & $0.240 \pm 0.013$ & $0.250 \pm 0.014$ & $0.246 \pm 0.009$ & $0.255 \pm 0.004$ \\
			& \textbf{RMSE }$\downarrow$ & $0.481 \pm 0.038$ & $0.438 \pm 0.033$ & $0.447 \pm 0.025$ & $0.444 \pm 0.016$ & $0.469 \pm 0.039$ & $0.517 \pm 0.042$ & $0.501 \pm 0.018$ & $0.528 \pm 0.033$ \\
			& $\mathbf{R^2}\uparrow$     & $0.926 \pm 0.012$ & $0.938 \pm 0.009$ & $0.936 \pm 0.007$ & $0.937 \pm 0.004$ & $0.929 \pm 0.012$ & $0.914 \pm 0.014$ & $0.920 \pm 0.006$ & $0.911 \pm 0.011$ \\
			
			\cline{1-10}
			
			\multirow{3}{*}{$M$-shot \gls{ICL}-\gls{LLM} \cite{HanICL}}
			& \textbf{MAE }$\downarrow$  & $0.269 \pm 0.022$ & $\mathbf{0.198 \pm 0.013}$ & $0.222 \pm 0.006$ & $0.207 \pm 0.014$ & $0.226 \pm 0.016$ & $0.242 \pm 0.009$ & $0.252 \pm 0.036$ & $0.235 \pm 0.012$ \\
			& \textbf{RMSE }$\downarrow$ & $0.527 \pm 0.063$ & $\mathbf{0.390 \pm 0.043}$ & $0.489 \pm 0.029$ & $0.437 \pm 0.060$ & $0.462 \pm 0.038$ & $0.550 \pm 0.045$ & $0.599 \pm 0.181$ & $0.524 \pm 0.054$ \\
			& $\mathbf{R^2}\uparrow$     & $0.910 \pm 0.021$ & $\mathbf{0.951 \pm 0.010}$ & $0.923 \pm 0.009$ & $0.938 \pm 0.016$ & $0.931 \pm 0.011$ & $0.903 \pm 0.016$ & $0.877 \pm 0.078$ & $0.912 \pm 0.019$ \\
			
			\hline

		\end{tabular}%
	}
\end{table*}
Finally, an important advantage of the \gls{CoT}-\gls{LLM} approach is its significantly improved prediction stability. In particular, the variance of the $R^2$-score is reduced by $65.36\%$ and $58.33\%$ for the downloading-while-driving and watching-Amazon-Prime-while-driving scenarios, respectively. This substantial reduction in variance highlights the robustness and reliability of the proposed \gls{CoT}-\gls{LLM} framework for mobile traffic prediction.
\subsection{Model Comparisons}
\label{subsec:Model}
In this section, we present the performance of the proposed \gls{CoT}-\gls{LLM} mobile traffic prediction framework across additional \glspl{LLM}. In particular, while the o4-mini model achieves acceptable performance, its weights are not publicly accessible, which limits its deployment to the OpenAI \gls{API}. This reliance can be both costly and time-consuming, since real-time traffic prediction may be affected by the communication and processing delays between the \gls{BS}/network provider and the OpenAI service. Thus, we evaluate the proposed approach using several open-weight models that can be deployed directly at the network provider. In particular, we consider the following models:
\begin{enumerate}
\item Ministral $3\ \ 3B$ \cite{mistral_ministral3_3b_docs_2025}.
\item Qwen $3 \ \ 8B$ \cite{qwen3}.
\item Phi $4$ reasoning $14 B$ \cite{abdin2025phi4reasoning}.
\end{enumerate}
These models are selected to reflect different model sizes and providers, which facilitates a more structured comparison across both dimensions. In particular, establishing performance baselines for one provider helps benchmark and interpret results when comparing models of varying sizes across different providers\footnote{Note that at the time this research is conducted these are the state of the art models}. \par 

In this paper, we use $M=2$ examples for all models to ensure a fair comparison, while the model-specific optimization of example selection is left for future work. Moreover, since the o4-mini \gls{LLM} is not open-weight, its exact number of parameters is not publicly disclosed; therefore, to avoid inaccurate approximations, we leave the \emph{Model Complexity} entry for o4-mini blank. Finally, each model relies on its own generated \emph{rationales}, i.e., the offline phase is performed separately for each model. \par 
\begin{figure*}[t]
    \centering

    \subfloat[\label{fig:myfig:a}]{
        \includegraphics[width=0.43\linewidth]{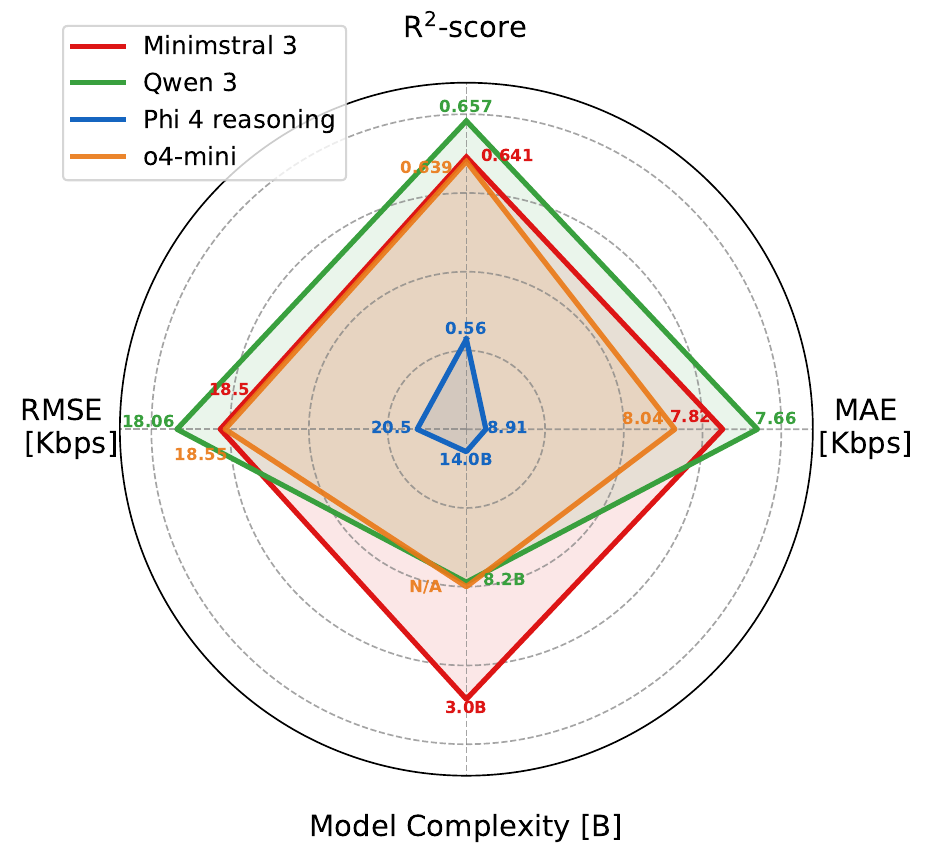}
    }
    \hfill
    \subfloat[\label{fig:myfig:b}]{
        \includegraphics[width=0.43\linewidth]{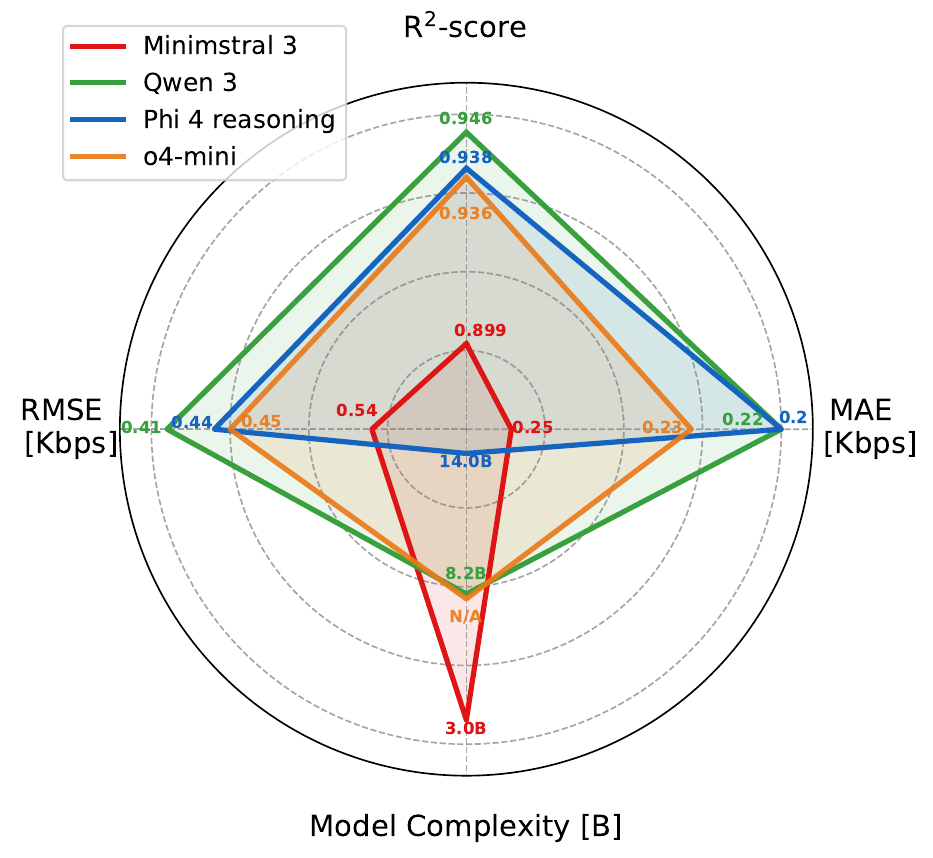}
    }

    \caption{Performance analysis of various \glspl{LLM} for2-shot \gls{CoT}-\gls{LLM} in (a) downloading while driving and (b) watching Amazon Prime while driving.}
    \label{fig:LLMs}
\end{figure*}
For ease of comparison, all four axes in Fig. \ref{fig:LLMs} are plotted so that a larger enclosed area corresponds to better overall performance. Specifically, $R^2$-score is plotted directly, where outward direction indicates higher values and better performance. On the other hand, \gls{MAE}, \gls{RMSE}, and \emph{Model Complexity} are inverted on their respective axes, where outward direction indicates lower error or fewer parameters, representing better performance. Overall, for each parameter, the axes are set up such that values closer to the outer circle represent better performance. For each metric, the figure highlights the average value over five independent runs. It can be observed that Qwen $3 \ 8B$ achieves the best overall performance across the considered settings, outperforming the o4-mini model. For instance, the Qwen $3 \ 8B$ attains \gls{MAE} of $7.66$, \gls{RMSE} of $18.05$, and $R^2$-score of $0.657$, whereas o4-mini model achieves an \gls{MAE} of $8.04$, \gls{RMSE} of $18.55$, and $R^2$-score of $0.639$. \par

Notably, the smallest model, Ministral $3 \ 3B$, achieves performance close to Qwen $3 \ 8B$, reaching $97.95\%$, $97.62\%$, and $97.56\%$, of Qwen $3 \ 8B$ in terms of \gls{MAE}, \gls{RMSE}, and $R^2$-score, respectively, while using only $36.59\%$ of the parameters. This indicates that strong performance can be obtained even with compact models. \par 
However, these results also show that increasing model size does not necessarily improve performance and may even lead to degradation. In the same setting, the Phi $4$ reasoning with $14B$ parameters achieves only $85.97\%$, $88.05\%$, and $85.23\%$ of Qwen $3 \ 8B$ in terms of \gls{MAE}, \gls{RMSE}, and $R^2$-score, respectively. Overall, the results suggest that medium-sized models can offer the best trade-off between accuracy and complexity, while smaller models can be deployed with only minor performance loss for a more cost-efficient approach. In contrast, deploying larger models such as Phi $4$ reasoning $14B$ may incur higher computational cost without providing additional performance gains. \par

\subsection{Complexity Analysis \& Deployment Considerations}
This subsection presents the computational complexity of the proposed \(M\)-shot \gls{CoT}-\gls{LLM} mobile traffic prediction framework. \par
During the offline phase, for a raw dataset with \(H\) seconds of traffic measurements, the overall complexity of constructing the historical windows (i.e., forming \(\left[\bm{\Gamma}, \mathbf{C}\right]\)) with \(K\) contextual features is \(\mathcal{O}\!\left(H \cdot N \cdot K\right)\), where \(N\) is defined in (\ref{eq:N}). Next, to generate the \emph{rationales} for each training example, we invoke the \gls{LLM} three times per sample, resulting in a total of \(3N\) \gls{LLM} calls. \par
In contrast, during the online phase, after establishing the processed training set and using the policy \(\pi\), the example-selection step requires \(\mathcal{O}\!\left(N \cdot M\right)\) operations for \(M\)-shot \gls{CoT}-\gls{LLM}. This is followed by a single \gls{LLM} call with  $\approx 854$ input and $\approx 152$ output tokens (output \emph{rationale} + answer for 2-shot) per test sample to produce the prediction, which is suitable for close to real-time, online operation. In practice, the traffic prediction and \gls{LLM} inference can be executed at the \glspl{BS} or on cloud servers, where such platforms are typically equipped with graphics processing units and artificial intelligence acceleration hardware to efficiently support the computational demands of \glspl{LLM}. Moreover, leveraging open-weight models, as demonstrated in the previous subsection, enables network providers to deploy \glspl{LLM} locally, thereby reducing both latency and the cost associated with external \glspl{API}. Under these infrastructure settings, an \gls{LLM} with fewer than \(15\) billion parameters is expected to achieve an inference time of roughly \(100\) ms \cite{ChittyVenkata2024LLMInferenceBench}, while the corresponding communication latency can be around \(50\) ms \cite{Fan2025LatencyAwareOffloadingEnergyControl}, depending on the hardware configuration and network conditions. These delays remain within acceptable limits for per-second traffic prediction. \par
In summary, the proposed \(M\)-shot \gls{CoT}-\gls{LLM} approach is feasible for real-world deployment in \gls{5G} and \gls{6G} networks, offering acceptable latency and manageable hardware requirements.

\section{CONCLUSION}
\label{sec:conc}
In this paper, we developed \gls{CoT}-enabled \gls{LLM} based mobile traffic prediction using real-world \gls{5G} measurements across different applications and mobility conditions. To address the limitations of standard \gls{ICL} prompting for numerical time-series prediction, we proposed a two-phase framework with an offline phase and an online phase. In the offline phase, we create structured \gls{CoT} demonstrations by generating rationales through a \gls{PCoT} pipeline. In the online phase, we use a lightweight example-selection policy to retrieve the most relevant demonstrations by comparing both the historical throughput trajectory and its short-term changes. This design allows effective few-shot inference while keeping the prompt length limited.\par 

Our results show that the proposed \gls{CoT}-\gls{LLM} approach can improve prediction accuracy and produce more stable outputs compared to \gls{ICL}-\gls{LLM} prompting and common baseline methods. In particular, the 2-shot \gls{CoT}-\gls{LLM} achieves improvements of up to $14.88\%$, $15.03\%$, and $22.41\%$ in \gls{MAE}, \gls{RMSE}, and $R^2$-score, respectively. We further show that optimizing the number of in-context examples yields additional gains across \gls{MAE}, \gls{RMSE}, and $R^2$-score. In addition, tests with multiple open-weight \glspl{LLM} indicate that locally deployable models can provide competitive performance, which reduces reliance on external \gls{API} services and supports close to real-time prediction. For future work, more advanced reasoning prompting strategies can be explored. 
\section*{Acknowledgment}

This work has been supported by MITACS, Ericsson Canada, and  Canada Research Chairs program.
\bibliographystyle{IEEEtran}
\bibliography{references}

@inproceedings{raca2020beyond,
  title     = {Beyond throughput, the next generation: A {5G} dataset with channel and context metrics},
  author    = {Raca, Darijo and Leahy, Dylan and Sreenan, Cormac J. and Quinlan, Jason J.},
  booktitle = {Proc. 11th ACM Multimedia Syst. Conf. (MMSys '20)},
  year      = {2020},
  pages     = {303--308},
  publisher = {ACM},
}

@inproceedings{dong2024survey,
  title={A survey on in-context learning},
  author={Dong, Qingxiu and Li, Lei and Dai, Damai and Zheng, Ce and Ma, Jingyuan and Li, Rui and Xia, Heming and Xu, Jingjing and Wu, Zhiyong and Chang, Baobao and others},
  booktitle={Proc. 2024 Conf. Empir. Methods Nat. Lang. Process.},
  pages={1107--1128},
  year={2024}
}

@ARTICLE{9352246,
  author={Yin, Xueyan and Wu, Genze and Wei, Jinze and Shen, Yanming and Qi, Heng and Yin, Baocai},
  journal={IEEE Trans. Intell. Transp. Syst.}, 
  title={Deep Learning on Traffic Prediction: Methods, Analysis, and Future Directions}, 
  year={2022},
  volume={23},
  number={6},
  pages={4927-4943},
  doi={10.1109/TITS.2021.3054840}}

@ARTICLE{11370176,
  author={Jiang, Feibo and Pan, Cunhua and Wang, Kezhi and Michiardi, Pietro and Dobre, Octavia A. and Debbah, Merouane},
  journal={IEEE J. Sel. Areas Commun.}, 
  title={From Large {AI} Models to Agentic {AI}: A Tutorial on Future Intelligent Communications}, 
  year={2026},
  volume={44},
  number={},
  pages={3507-3540},
  doi={10.1109/JSAC.2026.3660010}}

@inproceedings{NEURIPS2022_9d560961,
 author = {Wei, Jason and Wang, Xuezhi and Schuurmans, Dale and Bosma, Maarten and ichter, brian and Xia, Fei and Chi, Ed and Le, Quoc V and Zhou, Denny},
 booktitle = {Adv. Neural Inf. Process. Syst. (NeuroIPS)},
 pages = {24824-24837},
 title = {Chain-of-Thought Prompting Elicits Reasoning in Large Language Models},
 volume = {35},
 year = {2022}
}

@inproceedings{NEURIPS2022_8bb0d291,
 author = {Kojima, Takeshi and Gu, Shixiang and Reid, Machel and Matsuo, Yutaka and Iwasawa, Yusuke},
 booktitle = {Adv. Neural Inf. Process. Syst. (NeuroIPS)},
 pages = {22199--22213},
 title = {Large Language Models are Zero-Shot Reasoners},
 volume = {35},
 year = {2022}
}

@inproceedings{wang-etal-2023-plan,
    title = "Plan-and-Solve Prompting: Improving Zero-Shot Chain-of-Thought Reasoning by Large Language Models",
    author = "Wang, Lei  and
      Xu, Wanyu  and
      Lan, Yihuai  and
      Hu, Zhiqiang  and
      Lan, Yunshi  and
      Lee, Roy Ka-Wei  and
      Lim, Ee-Peng",
    booktitle = "Proc. 61st Annu. Meeting Assoc. Comput. Linguistics",
    year = "2023",
    address = "Toronto, Canada",
    pages = "2609--2634",
}

@misc{openai2025o4mini,
  author = {OpenAI},
  title = {OpenAI o3 and o4-mini System Card},
  year = {2025},
}

@article{HanICL,
  author  = {Zhang, Han and Bin Sediq, Akram and Afana, Ali and Erol-Kantarci, Melike},
  title   = {Mobile Traffic Prediction Using {LLMs} With Efficient In-Context Demonstration Selection},
  journal = {IEEE Transactions on Communications},
  year    = {2025},
  volume  = {73},
  number  = {11},
  pages   = {11170--11185}
}

@article{10.1007/s10922-016-9392-x,
author = {Dalmazo, Bruno L. and Vilela, Jo\~{a}o P. and Curado, Marilia},
title = {Performance Analysis of Network Traffic Predictors in the Cloud},
year = {2017},
issue_date = {April     2017},
publisher = {Plenum Press},
address = {USA},
volume = {25},
number = {2},
journal = {J. Netw. Syst. Manage.},
month = apr,
pages = {290–320},
numpages = {31},
}

@article{TianL21-1,
  title = {Network traffic prediction method based on autoregressive integrated moving average and adaptive Volterra filter},
  author = {Zhongda Tian and Feihong Li},
  year = {2021},
  doi = {10.1002/dac.4891},
  url = {https://doi.org/10.1002/dac.4891},
  researchr = {https://researchr.org/publication/TianL21-1},
  cites = {0},
  citedby = {0},
  journal = {Int. J. Commun. Sys.},
  volume = {34},
  number = {12},
}

@misc{mistral_ministral3_3b_docs_2025,
  title        = {Ministral 3 3B},
  author       = {{Mistral AI}},
  year         = {2025},
  month        = dec,
  howpublished = {Mistral Docs (Open v25.12)},
  note         = {Accessed 2025-12-28}
}

@article{abdin2025phi4reasoning,
  title         = {Phi-4-reasoning Technical Report},
  author        = {Marah Abdin and others},
  journal       = {arXiv preprint arXiv:2504.21318},
  year          = {2025},
}

@article{qwen3,
  title   = {Qwen3 Technical Report},
  author  = {An Yang and others},
  journal = {arXiv preprint arXiv:2505.09388},
  year    = {2025},
}

@inproceedings{ChittyVenkata2024LLMInferenceBench,
  author    = {Chitty-Venkata, Krishna and others},
  title     = {{LLM}-Inference-Bench: Inference Benchmarking of Large Language Models on {AI} Accelerators},
  booktitle = {Proc. SC24-W: Workshops Int. Conf. High Perform. Comput., Netw., Storage Anal.},
  year      = {2024},
  address   = {Atlanta, GA, USA},
  pages     = {1362--1379},
}

@article{Fan2025LatencyAwareOffloadingEnergyControl,
  author  = {Fan, Weibei and Xiao, Fu and Pan, Yao and Chen, Xiaobai and Han, Lei and Yu, Shui},
  title   = {Latency-aware Joint Task Offloading and Energy Control for Cooperative Mobile Edge Computing},
  journal = {IEEE Trans. Serv. Comput.},
  year    = {2025},
  volume  = {18},
  number  = {3},
  pages   = {1515--1528},
}

@article{Mei2022Realtime,
	author  = {Mei, Lifan and Gou, Jinrui and Cai, Yujin and Cao, Houwei and Liu, Yong},
	title   = {Realtime mobile bandwidth and handoff predictions in {4G}/{5G} networks},
	journal = {Comput. Netw.},
	volume  = {204},
	pages   = {108736},
	year    = {2022},
	month   = feb,

}

@article{Saad2020SixG,
	author  = {Walid Saad and Mehdi Bennis and Mingzhe Chen},
	title   = {A Vision of {6G} Wireless Systems: Applications, Trends, Technologies, and Open Research Problems},
	journal = {IEEE Commun. Mag.},
	volume  = {58},
	number  = {9},
	pages   = {74--80},
	year    = {2020},
}

@article{Lykakis2025DataTrafficPrediction,
	author    = {Evangelos Lykakis and Ioannis O. Vardiambasis and Evangelos Kokkinos},
	title     = {Data Traffic Prediction for {5G} and Beyond: Emerging Trends, Challenges, and Future Directions: A Scoping Review},
	journal   = {Electronics},
	volume    = {14},
	number    = {23},
	pages     = {4611},
	year      = {2025},
}

@ARTICLE{9112608,
  author={Tedjopurnomo, David Alexander and Bao, Zhifeng and Zheng, Baihua and Choudhury, Farhana Murtaza and Qin, A. K.},
  journal={IEEE Trans. Knowl. Data Eng.}, 
  title={A Survey on Modern Deep Neural Network for Traffic Prediction: Trends, Methods and Challenges}, 
  year={2022},
  volume={34},
  number={4},
  pages={1544-1561},
  }

@INPROCEEDINGS{10279008,
  author={Iturria-Rivera, Pedro Enrique and Chenier, Marcel and Herscovici, Bernard and Kantarci, Burak and Erol-Kantarci, Melike},
  booktitle={Proc. IEEE Int. Conf. Commun. (ICC) 2023}, 
  title={RL meets Multi-Link Operation in IEEE 802.11be: Multi-Headed Recurrent Soft-Actor Critic-based Traffic Allocation}, 
  year={2023},
  volume={},
  number={},
  pages={4001-4006},
 }

@article{Brown2020LanguageModelsFewShot,
	author    = {Tom B. Brown and Benjamin Mann and Nick Ryder and Melanie Subbiah and Jared Kaplan and others},
	title     = {Language Models are Few-Shot Learners},
	journal   = {Adv. Neural Inf. Process. Syst. (NeuroIPS)},
	volume    = {33},
	pages     = {1877--1901},
	year      = {2020}
}

@inproceedings{Trinh2018MobileTrafficLSTM,
	author    = {Trinh, Hoang Duy and Giupponi, Lorenza and Dini, Paolo},
	title     = {Mobile Traffic Prediction from Raw Data Using {LSTM} Networks},
	booktitle = {Proc. IEEE Int. Symp. Personal, Indoor and Mobile Radio Commun. (PIMRC)},
	year      = {2018},
	publisher = {IEEE},
	address   = {Bologna, Italy},
	pages     = {1--6},
}

@inproceedings{Bai2020AdaptiveGraphCRN,
	author    = {Bai, Lei and Yao, Lina and Li, Can and Wang, Xianzhi and Wang, Can},
	title     = {Adaptive Graph Convolutional Recurrent Network for Traffic Forecasting},
	booktitle = {Proc. Annual Conf. Neural Inf. Process. Syst. (NeurIPS)},
	year      = {2020},
}

@ARTICLE{9673775,
	author={Fang, Yini and Ergüt, Salih and Patras, Paul},
	journal={IEEE Commun. Lett.}, 
	title={{SDGNet}: A Handover-Aware Spatiotemporal Graph Neural Network for Mobile Traffic Forecasting}, 
	year={2022},
	volume={26},
	number={3},
	pages={582-586},
}

@inproceedings{Vaswani2017Attention,
	author    = {Vaswani, Ashish and Shazeer, Noam and Parmar, Niki and Uszkoreit, Jakob and Jones, Llion and Gomez, Aidan N. and Kaiser, {\L}ukasz and Polosukhin, Illia},
	title     = {Attention Is All You Need},
	booktitle = {Adv. Neural Inf. Process. Syst. (NeuroIPS)},
	year      = {2017},
	pages     = {5998--6008},

}

@ARTICLE{10946853,
	
	author={Kougioumtzidis, Georgios and Poulkov, Vladimir K. and Lazaridis, Pavlos I. and Zaharis, Zaharias D.},
	
	journal={IEEE Trans. Artif. Intell.}, 
	
	title={Mobile Network Traffic Prediction Using Temporal Fusion Transformer}, 
	
	year={2025},
	
	volume={6},
	
	number={10},
	
	pages={2685-2699},
}

@ARTICLE{9919315,
	author={Hu, Yahui and Zhou, Yujiang and Song, Junping and Xu, Luyang and Zhou, Xu},
	journal={IEEE Trans. Netw. Serv. Manage.}, 
	title={Citywide Mobile Traffic Forecasting Using Spatial-Temporal Downsampling Transformer Neural Networks}, 
	year={2023},
	volume={20},
	number={1},
	pages={152-165},
}

@ARTICLE{10840287,
	author={Gong, Jiahui and Liu, Yu and Li, Tong and Ding, Jingtao and Wang, Zhaocheng and Jin, Depeng},
	journal={IEEE Trans. Mobile Comput.}, 
	title={{STTF}: A Spatiotemporal Transformer Framework for Multi-task Mobile Network Prediction}, 
	year={2025},
	volume={24},
	number={5},
	pages={4072-4085},
	}

@ARTICLE{10114636,
	author={Gu, Bo and Zhan, Junhui and Gong, Shimin and Liu, Wanquan and Su, Zhou and Guizani, Mohsen},
	journal={IEEE Trans. Wireless Commun.}, 
	title={A Spatial-Temporal Transformer Network for City-Level Cellular Traffic Analysis and Prediction}, 
	year={2023},
	volume={22},
	number={12},
	pages={9412-9423},
}

@article{Zhou2024LLMTelecomSurvey,
	author    = {Zhou, Hao and Hu, Chengming and Yuan, Ye and Cui, Yufei and Jin, Yili and Chen, Can and Wu, Haolun and Yuan, Dun and Jiang, Li and Wu, Di and Liu, Xue and Zhang, Charlie and Wang, Xianbin and Liu, Jiangchuan},
	title     = {Large Language Model ({LLM}) for Telecommunications: A Comprehensive Survey on Principles, Key Techniques, and Opportunities},
	journal   = {IEEE Commun. Surveys Tuts.},
	year      = {2024},
	volume    = {27},
	number    = {3},
	pages     = {1955--2005},
}

@inproceedings{Zhang2024LLMWirelessIntrusion,
	author    = {Zhang, Han and Sediq, Akram Bin and Afana, Ali and Erol-Kantarci, Melike},
	title     = {Large Language Models in Wireless Application Design: In-Context Learning-Enhanced Automatic Network Intrusion Detection},
	booktitle = {Proc. IEEE Global Commun. Conf. (GLOBECOM)},
	year      = {2024},
	pages     = {2479--2484},
}

@inproceedings{Habib2025LLMIntentNetOpt,
	author    = {Habib, Md Arafat and Iturria Rivera, Pedro Enrique and Ozcan, Yigit and Elsayed, Medhat H. M. and Bavand, Majid and Gaigalas, Raimundus and Erol-Kantarci, Melike},
	title     = {{LLM}-Based Intent Processing and Network Optimization Using Attention-Based Hierarchical Reinforcement Learning},
	booktitle = {Proc. 2025 IEEE Wireless Commun. Netw. Conf. (WCNC)},
	year      = {2025},
	pages     = {1--6}
}

@article{Hu2025SelfRefinedTrafficLLM,
	author  = {Hu, Chengming and Zhou, Hao and Wu, Di and Chen, Xi and Yan, Jun and Liu, Xue},
	title   = {Self-Refined Generative Foundation Models for Wireless Traffic Prediction},
	journal = {IEEE Trans. Veh. Technol.},
	year    = {2025},
}

@article{Chang2025LLM4TS,
	author  = {Chang, Ching and Wang, Wei-Yao and Peng, Wen-Chih and Chen, Tien-Fu},
	title   = {{LLM4TS}: Aligning Pre-Trained {LLMs} as Data-Efficient Time-Series Forecasters},
	journal = {ACM Trans. Intell. Syst. Technol.},
	year    = {2025},
	volume  = {16},
	number  = {3},
	pages   = {1--20},
}

@inproceedings{Cao2024TEMPO,
	author    = {Cao, Defu and Jia, Furong and Arik, Sercan O. and Pfister, Tomas and Zheng, Yixiang and Ye, Wen and Liu, Yan},
	title     = {TEMPO: Prompt-Based Generative Pre-Trained Transformer for Time Series Forecasting},
	booktitle = {Proc. Int. Conf. Learn. Represent. (ICLR)},
	year      = {2024},
}

@inproceedings{Huang2025ReasoningAI6G,
	author    = {Huang, Liming and Wu, Yulei and Simeonidou, Dimitra},
	title     = {Reasoning {AI} Performance Degradation in {6G} Networks with Large Language Models},
	booktitle = {Proc. 2025 IEEE Wireless Commun. Netw. Conf. (WCNC)},
	year      = {2025},
	pages     = {1--6},
}

@article{wang2025chain,
	title={Chain-of-Thought for Large Language Model-empowered Wireless Communications},
	author={Wang, Xudong and Zhu, Jian and Zhang, Ruichen and Feng, Lei and Niyato, Dusit and Wang, Jiacheng and Du, Hongyang and Mao, Shiwen and Han, Zhu},
	journal={arXiv preprint arXiv:2505.22320},
	year={2025}
}

@article{Bandt2002PermutationEntropy,
  author  = {Bandt, Christoph and Pompe, Bernd},
  title   = {Permutation Entropy: A Natural Complexity Measure for Time Series},
  journal = {Physical Review Letters},
  volume  = {88},
  number  = {17},
  pages   = {174102},
  year    = {2002},
}

\end{document}